\documentclass[11pt]{article}
\usepackage{jcappub,natbib,float,caption}

\usepackage{amssymb}
\usepackage{amsmath}
\usepackage{hyperref}

\usepackage{bm,comment}
\usepackage{graphicx}
\usepackage{graphics}
\usepackage{dcolumn}
\usepackage{color}
\usepackage{rotate}
\usepackage{fancyhdr} 

\def\be{\begin{equation}}
\def\ee{\end{equation}}
\def\bea{\begin{eqnarray}}
\def\eea{\end{eqnarray}}
\def\nn{\nonumber}

\def\coH{\mathcal{H}}
\def\comm#1#2{\left[#1,\,#2\right]}

\newcommand{\refeq}[1]{Eq.~(\ref{eq:#1})}          
\newcommand{\refeqs}[2]{Eqs.~(\ref{eq:#1})--(\ref{eq:#2})}          
\newcommand{\reffig}[1]{Fig.~\ref{fig:#1}}          
\newcommand{\refsec}[1]{Sec.~\ref{sec:#1}}
\newcommand{\refapp}[1]{App.~\ref{app:#1}}

\newcommand{\chg}[1]{#1}

\def\Ga#1#2#3{\Gamma^{#1}_{#2#3}}
\def\pdGa#1#2#3#4{\Gamma^{#1}_{#2#3,#4}}

\def\Cten#1#2#3{C^{#1}{}_{#2#3}}
\def\tilGa#1#2#3{\tilde{\Gamma}^{#1}_{#2#3}}
\def\pdtilGa#1#2#3#4{\tilde{\Gamma}^{#1}_{#2#3,#4}}
\def\pdCten#1#2#3#4{C^{#1}{}_{#2#3,#4}}

\renewcommand{\v}[1]{\mathbf{#1}}
\newcommand{\vx}{\v{x}}
\def\L{\Lambda}
\def\O{\mathcal{O}}
\def\eps{\varepsilon}
\def\ba#1\ea{\begin{align}#1\end{align}}
\newcommand{\vs}{\nonumber\\}

\title{Conformal Fermi Coordinates}

\author[a]{Liang Dai,}

\author[b]{Enrico Pajer}

\author[c]{and Fabian Schmidt}

\affiliation[a]{Department of Physics and Astronomy, Johns Hopkins University, 3400 N. Charles St., Baltimore, MD 21218, USA}

\affiliation[b]{Institute for Theoretical Physics and Center for Extreme Matter and Emergent Phenomena,
Utrecht University, Leuvenlaan 4, 3584 CE Utrecht, The Netherlands}

\affiliation[c]{Max-Planck-Institut f\"ur Astrophysik, Karl-Schwarzschild-Str. 1, 85741 Garching, Germany}

\abstract{
Fermi Normal Coordinates (FNC) are a useful frame for isolating the
locally observable, physical effects of a long-wavelength spacetime
perturbation.  Their cosmological application, however, is hampered
by the fact that they are only valid on scales much smaller than the 
horizon.  We introduce a generalization that we call 
\emph{Conformal Fermi Coordinates} (CFC).  CFC preserve all
the advantages of FNC, but in addition are valid outside the horizon.  
They allow us to calculate the coupling of long-
and short-wavelength modes on all scales larger than the sound horizon
of the cosmological fluid, starting from the epoch of inflation until
today, by removing the complications of the second order
Einstein equations to a large extent, and eliminating all gauge ambiguities.  
As an application, we present a calculation
of the effect of long-wavelength tensor modes on small scale density
fluctuations.  We recover previous results, but clarify the physical
content of the individual contributions in terms of locally measurable
effects and ``projection'' terms.
}

\begin{document}

\maketitle
\flushbottom

\section{Introduction}
\label{sec:intro}

When performing a calculation one may choose any set of coordinates. However, choose the ``wrong'' coordinates, and you will be sorry.\footnote{This quote is allegedly due to S.~Weinberg.} This is especially true in general relativity and its application to cosmological perturbation theory. A first and most obvious issue is that the ``wrong'' coordinates might substantially increase the computational complexity of the problem. A second and maybe more subtle issue is the risk of incorrectly interpreting a legitimate solution of the equations of motion in terms of measurements because the coordinates in which that solution has been derived are different from the natural coordinates of the observer. One often loosely describes this situation as \textit{gauge artifacts}. Of course, the computation of a physical observable, defined in an explicit operational sense, must give the same result in any set of coordinates. However, even the simplest operational measurement might become very subtle to define when using the wrong set of coordinates. In the context of cosmology, Maldacena's consistency relation \cite{Maldacena:2002vr,Creminelli:2004yq} provides a sharp example of how subtle gauge artifacts can be. The consistency relation states that, for single clock inflation, primordial non-Gaussianity is of the local form with $ f_{NL}^{\rm loc} \propto (n_{s}-1) $, where $ n_{s} $ is the spectral tilt of the primordial scalar power spectrum. One might naively think that the locally measured short-scale scalar power spectrum in the presence of a long scalar mode (say of $ \zeta $) is different from what one would measure locally in the absence of the long mode. One might even be tempted to speculate that if $ (n_{s}-1) $ were allowed to be somewhat larger than in standard slow-roll inflation, one could have a significant locally observable effect. This interpretation and speculation are both incorrect because the consistency condition is derived using the comoving coordinates of an unperturbed universe, while the ruler of a local observer would instead change in the presence of the long mode in precisely such a way that she would not see any difference from an unperturbed universe (see \cite{Pajer:2013ana} for a detailed discussion).  In fact, the consistency relations discussed widely
in the literature \cite{Maldacena:2002vr,Creminelli:2004yq,Kehagias:2013paa,Cheung:2007sv,Creminelli:2012ed,
Hinterbichler:2012nm,Senatore:2012wy,Hinterbichler:2013dpa,Goldberger:2013rsa,Chen:2013aj} encode
the gauge freedom or diffeomorphism invariance remaining in the description of single-field slow roll
inflation.  Single-clock inflation does lead to some observable non-Gaussianity of the local shape in say the Cosmic Microwave Background (CMB) and Large Scale Structure.  However, this is due to the effect of large-scale perturbations (within the current horizon) on the propagation of photons from the source to the Earth (``projection effects'') which are entirely independent of interactions during inflation and do not scale as $n_s-1$.

Choosing the natural set of coordinates of a local observer is very convenient because they are directly related to local measurements, hence explicitly eliminating any gauge artifacts. In most applications to cosmology, the local observer is in an inertial frame, free-falling in the local gravitational potential.  Such an inertial observer can describe the neighborhood around her worldline as flat spacetime with corrections that grow with the square of the distance from the observer's geodesic times second derivatives of the spacetime metric as described by the Riemann tensor.  This holds in an arbitrary spacetime, and the set of coordinates is known as Fermi Normal Coordinates (FNC) \cite{Manasse:1963zz} and it is implicitly used by all laboratory experiments for which the gravitational field of the Earth is negligible, such as for example at the Large Hadron Collider. A major drawback that hampers the usage of FNC in cosmology is the fact that the Friedmann-Lemaitre-Robertson-Walker (FLRW) metric has second derivatives of order $H^2$, where $H$ is the Hubble parameter. In such a spacetime, FNC cover at most a patch of size $ H^{-1} $ which is too small to describe horizon-scale or super-horizon modes $ k\lesssim aH $. Hence one cannot use FNC to connect primordial perturbations generated during the early universe to the late time evolution leading up to the time of observation. 

In this paper we formalize the construction of a generalization of FNC, which we call \textit{Conformal Fermi Coordinates} (CFC).\footnote{They were first introduced in a less formal way in \cite{Pajer:2013ana}.  We make the connection to their construction in \refsec{perturbed-FLRW}.} These are the coordinates of a local observer that describes the spacetime in a neighborhood of her worldline as an FLRW spacetime.  The corrections from the unperturbed FLRW spacetime again grow quadratically in the distance from the worldline. However, depending on the structure of the spacetime, the corrections can stay small on scales much larger than the Hubble horizon.  Thus, these coordinates share all the advantages of FNC but are valid on super-horizon scales.  Indeed, \textit{just as particle physicists implicitly use FNC, as cosmologists we implicitly use CFC in describing our background universe}, allowing us to ignore deviations from the FLRW spacetime that are beyond our current horizon (even though they could be very large).  

The perks of using CFC in cosmology are best described for a toy spacetime that consists of just one long- and one short-wavelength perturbation on top of a homogenous and isotropic background. These perturbations might be scalars, vectors, tensors or a combination of them.  The following considerations then apply:

\begin{itemize}
\item The CFC defined for an observer that is inertial with respect to the long mode can be used to describe the dynamics of the short mode. At the observer's location, it describes a local FLRW universe with tidal corrections, whether either or both the long and the short mode are inside or outside the Hubble horizon, as long as the long mode is outside the sound horizon of the fluid. For instance, in a universe dominated by dark matter, a spherical overdense region locally collapses in exactly the same way as an over-critical universe. Regardless of the size of the region compared to the horizon, the picture applies as long as the region is larger than the free-streaming distance of the dark matter.

\item As it is the case for FNC, results computed in CFC have a very transparent physical interpretation: They correspond precisely to what a local observer would measure, as she would find herself living in a FLRW spacetime whose Hubble parameter matches the locally-measured rate of expansion and whose spatial curvature matches the local spatial geometry.  In this sense, CFC are the complement of consistency relations, in that they remove the diffeomorphim invariance to isolate all locally observable effects.

\item Using CFC and given some primordial perturbations, one can organize the computation of observables in a convenient and physically transparent way (see e.g. \cite{Pajer:2013ana}) by transforming to CFC at some point during inflation and then following the evolution of small scales in this frame.  In a further step, one computes the so-called projection effects to relate local physics to what a distant observer sees.  This is done perhaps most simply by transforming to FNC first (which takes into account the local CFC scale factor) and then using the ruler perturbations of \cite{Schmidt:2012ne} (\refsec{relation-FNC}).

\item The computation of non-linear effects that couple long and short modes is substantially simplified.  This in particular applies to the case where the
short-wavelength modes are far inside the horizon (which is the case typically of practical interest).  We show here that \emph{it is sufficient to include the much simpler non-linearities in the fluid equations while keeping the Einstein equations to linear order} in the small-scale modes.  This had already been used in \cite{Schmidt:2013gwa}.   We also show how the CFC calculation can be extended to the case in which the short mode is super-horizon.  In this regime one in general needs to include certain non-linear terms in the Einstein equations, and some subtleties arise in dealing with spatial derivative operators.  In this regime, which is typically not phenomenologically interesting, the calculation is not any simpler than a global calculation, but it does retain the advantage of the more transparent interpretation.  
\end{itemize}

The remainder of the paper is organized as follows. In Sec. \ref{sec:FNC} we set the stage with a brief review of FNC. We then describe the construction of the CFC in Sec. \ref{sec:CFC-main}. In Sec. \ref{sec:perturbed-FLRW} we discuss how to choose a local scale factor appearing in CFC for a perturbed FLRW universe, and give formulae for the conversion to CFC.  We also discuss the subtle differences to \cite{Pajer:2013ana}.  In Sec. \ref{sec:EE} we derive the relevant terms in the Einstein and fluid equations needed to capture the leading-order couplings between long and short modes.  This provides one of the main motivations for using CFC.  We discuss in detail the relation to observations made by a distant observer in \refsec{relation-FNC}.  Sec. \ref{sec:long-tensor} discusses application of the CFC to a long tensor mode affecting short scalar modes, extending the results of \cite{Schmidt:2013gwa} to the case in which the short mode is super-horizon, and clarifying the interpretation of their result.  We finally conclude in \refsec{concl}. Useful mathematical results are collected in the appendices.  A more detailed treatment of long-wavelength scalar perturbations will be presented in a companion paper~\cite{CFCpaper2}. 

\section{Recap of Fermi Normal Coordinates (FNC)}
\label{sec:FNC}

We first briefly review the basic concept of FNC and its construction before generalizing to CFC in the context of cosmology. 

Consider a free-falling observer, whose trajectory is a time-like geodesic, which we call the central geodesic.  One can choose that it defines the spatial origin of the coordinate system at all times, and further that the tangent vector to the geodesic defines the time direction.  The most obvious choice is simply to take the proper time $t_F$ along the central geodesic as time coordinate.  We then choose an orthonormal set of spatial basis vectors, completing the tetrad $(e_a)^\mu$, such that at some point on the geodesic $g^F_{\mu\nu}=\eta_{\mu\nu}$.  
Parallel transport of these basis vectors
ensures that this is valid for any point along the geodesic.  Further,
the fact that the observer is free-falling allows us to choose coordinates so that for any point along the central geodesic $(\partial_\rho g_{\mu\nu})_F=0$. Thus, all Christoffel connections vanish, $(\Ga{\rho}{\mu}{\nu})_F$. For points off the central geodesic, the FNC metric differs from the flat one only at quadratic order in the spatial FNC coordinate $x^i_F$.  In other words, the FNC are the natural coordinates a free-falling observer would use to describe local measurements.  The leading gravitational effect she would observe is a tidal field.  

Since the FNC is only necessarily flat in the vicinity of the central geodesic, the parameterization of the spatial coordinates is not unique (see e.g.~\cite{Pajer:2013ana}). However, a conceptually simple geometrical construction of FNC satisfying all of the aforementioned requirements exists~\cite{Manasse:1963zz}. 
The slice of simultaneity containing a given point $P$ on the central geodesic is defined as the surface spanned by spatial geodesics that radiate outward from $P$. The magnitude of the FNC position $x^i_F$ can be fixed by the {\it proper} length from $P$ along the spatial geodesic. To be more specific, the spatial geodesic connecting $P$ (having FNC coordinates $ x^\mu_F(P)=\{\tau_F,\bm{0}\}$) and another point $Q$ (having FNC coordinates $x^\mu_F(Q)=\{\tau_F,x^i_F\}$) on that slice can be parameterised by an affine parameter $\lambda$,
where $\lambda=0$ at $P$ and $\lambda=1$ at $Q$. The direction of the geodesic is fixed by the initial condition
\bea
\label{eq:FNC-alpha1}
\left. \frac{dx^\mu}{d\lambda} \right|_{\lambda=0} = x^i_F (e_i)^\mu_P,
\eea  
where $x^i_F$ is taken to be the FNC position for $Q$. This ensures that the proper distance squared is given by $\delta_{ij} x^i_F x^j_F$ in the vicinity of $P$, since $g_{\mu\nu} (e_i)^\mu_P (e_j)^\nu_P = \delta_{ij}$. Under this construction, the coordinate transformation from some coordinate system $x^\mu$ to the FNC coordinate $x^\mu_F$ can be computed order-by-order in $x^i_F$.  One repeatedly uses the geodesic equation to determine the coefficients of the series expansion $x^\mu(\lambda)$ in $\lambda$, which are completely determined by derivatives of the metric evaluated at $P$. Truncation at third order is sufficient to determine the transformation Jacobian matrix $\partial x^\mu / \partial x^\nu_F$ to second order in $x^i_F$ and the tidal corrections in the FNC metric.

The FNC metric can then be found order-by-order in $x^i_F$ using\footnote{Note that the coordinate transformation enters not only through $\partial x^\alpha / \partial x^\mu_F$, but also through the explicit coordinate shift from the central geodesic, $g_{\alpha\beta}(x) = \left( g_{\alpha\beta} \right)_P + \left( g_{\alpha\beta,\mu} \right)_P \delta x^\mu + \cdots$, where $\delta x^\mu$ starts at linear order in $x^i_F$.}
\bea
\label{eq:metric-transform}
g^F_{\mu\nu}(x_F) = \frac{\partial x^\alpha}{\partial x^\mu_F} \frac{\partial x^\beta}{\partial x^\nu_F} g_{\alpha\beta}(x)\,.
\eea
Here and throughout, a sub- or superscript $F$ denotes quantities in the FNC or CFC frame.  
One can then verify that this construction successfully realizes that the FNC metric is flat up to second order in the spatial deviation from the central geodesic,
\bea
g^F_{\mu\nu}(x_F) = \eta_{\mu\nu} + \mathcal{O}[R^{F}_{\mu\nu\alpha\beta}(x^i_F)^2]\,,
\eea
where the magnitude of the corrections is given by the Riemann tensor (transformed to the FNC frame).  
As an example, we can consider a Friedmann-Robertson-Walker spacetime, given
by
\be
ds^2 = a^2(\tau)\left[-d\tau^2 + \left(1 + \frac14 K \vx^2\right)^{-2} \delta_{ij} dx^i dx^j \right]\,,
\label{eq:dsFLRW}
\ee
where $\vx^2 = \delta_{ij} x^i x^j$ and $K$ is the curvature constant.  This
metric becomes in FNC (e.g., \cite{Baldauf:2011bh})\footnote{There is residual gauge freedom in the spatial metric components as discussed in \cite{Pajer:2013ana,CFCpaper2}.}
\ba
g^F_{00} =\:& -1 + \left(\frac{dH}{dt_F} + H^2(t_F)\right) \vx_F^2\,;
\quad g^F_{0i} = 0\,; \vs
g^F_{ij} =\:& \left[ 1 - \frac12\left(H^2(t_F) + \frac{K}{a^2(t_F)} \right)\vx_F^2 \right] \delta_{ij}  \,,
\label{eq:FNCFLRW}
\ea
where $H = a^{-1} da/dt = \coH/a$.  
Of course, one can derive a more general expression allowing for a perturbed
FLRW metric.  However, it is already clear from \refeq{FNCFLRW} that the 
FNC in the cosmological context are only valid on scales that are much smaller
than the horizon, since we are expanding perturbatively in 
$H x^i_F$; if this quantity becomes order one, the perturbative description
of the FNC metric breaks down.  We will now show how the conformal generalization of FNC can get around this limitation.
 
\section{Conformal Fermi Coordinates (CFC)}
\label{sec:CFC-main}

We now describe the construction of the conformal generalization of FNC.  
Since we shall not discuss FNC and CFC at the same time in the following, 
the subscript $F$ will be reserved for CFC hereafter.  
The CFC are constructed in the vicinity of a timelike central geodesic, just
like the ordinary FNC. However, unlike the FNC, it does not restrict the local spacetime to be Minkowski, but allows for a homogeneous expansion over time;  that is, the lowest order CFC metric is an FLRW spacetime \refeq{dsFLRW}.  Thus, the CFC metric takes the following form
\bea
\label{eq:CFC-requirement}
g^F_{\mu\nu}(x^\mu_F) = a^2_F(\tau_F) \left[ - \eta_{\mu\nu} + h^F_{\mu\nu} (\tau_F, x^i_F) \right],\qquad h^F_{\mu\nu} = \mathcal{O}[(x^i_F)^2] .
\eea 
Like in the case of FNC, corrections to the conformally flat part start at quadratic order in $x^i_F$. However, a few additional subtleties arise. First, the CFC time $\tau_F$ should be some suitable conformal time rather than the observer's proper time. Besides, a suitable local scale factor $a_F(\tau_F)$ should be defined in a physical way.  In particular, if $g_{\mu\nu}$ describes an unperturbed FLRW metric (but given, for example, in some unusual coordinates), then the CFC construction should yield the metric in the canonical FLRW form, \refeq{dsFLRW}.  These considerations motivate our generalization of the usual FNC construction presented in the previous section.

\subsection{Constructing CFC}
\label{sec:define-CFC}

\begin{figure}[t!]
\centering
\includegraphics[scale=0.4]{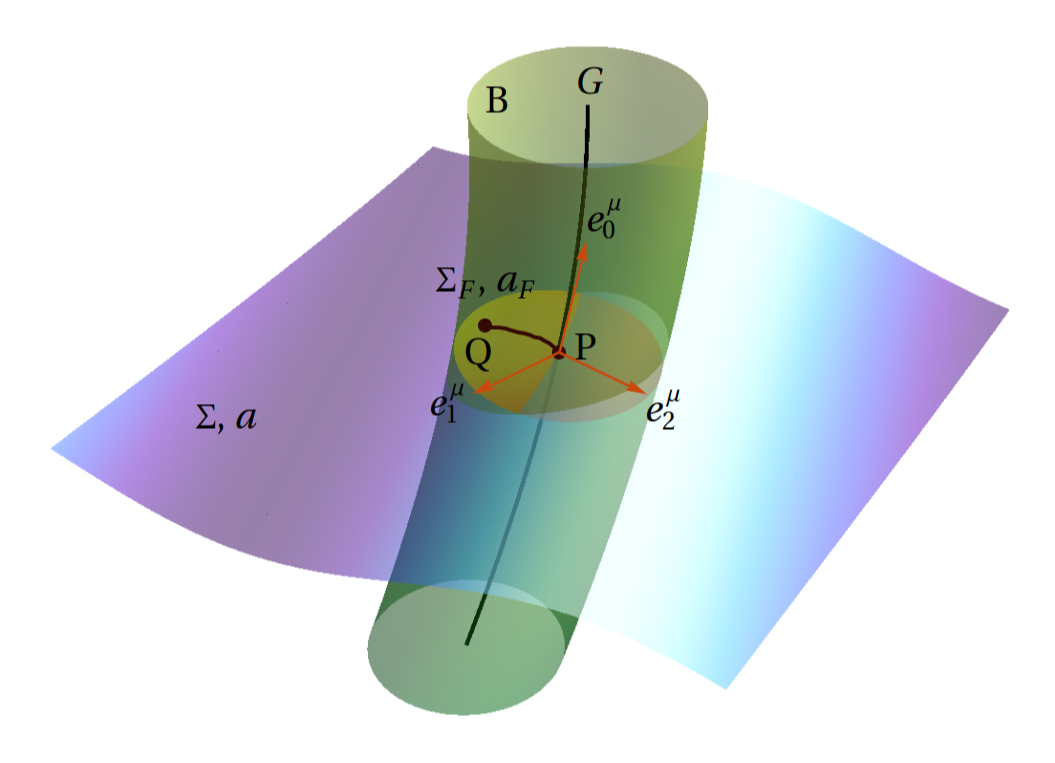}
\caption{Construction of the CFC. At point $P$, the observer's geodesic $G$ intersects a spatial hypersurface $\Sigma$, having constant conformal time $\tau$ and scale factor $a(\tau)$ in some global coordinate system. The spatial hyper-surface $\Sigma_F$, having constant CFC conformal time $\tau_F$ and CFC scale factor $a_F(\tau_F)$, also intersects $G$ at $P$, but does not coincide with $\Sigma$ in general. Another point $Q$ on $\Sigma_F$ is connected to $P$ by a conformal geodesic, and is parameterized by the same $\tau_F$ but nonzero $x^i_F$. The CFC coordinates are valid within a tubelike region bounded by hyper-surface $B$ surrounding $G$.\label{fig:FigureIllustration}
}
\end{figure}

The geometrical relation between the CFC and the global coordinates is sketched in \reffig{FigureIllustration}.  Throughout this paper, ``global coordinates'' refers to some set of coordinates valid at least in the region surrounding the geodesic considered (for example in one of the standard gauges of cosmological perturbation theory).  This name is chosen to distinguish it from the local construction of the CFC and FNC.

First, we choose the same set of orthonomal tetrads $(e^\mu_0)_P$ and $(e^\mu_i)_P,\,i=1,2,3$ as in the construction of FNC. We also parameterize the observer's geodesic, which is also the spatial origin of CFC, in terms of the proper time $t_F$ in the usual way.  

Next, consider a spacetime scalar $a_F(x)$, which we require to be positive at least in a finite region around the central geodesic.  We can then define a ``conformal proper time'' $\tau_F$ through
\be
d\tau_F = a_F^{-1}(P(t_F)) dt_F\,,
\ee
where $P(t_F)$ is the point along the central geodesic at proper time $t_F$.  
This can be integrated to yield a unique relation $\tau_F(t_F)$ (up to
an integration constant which can be absorbed into a redifinition of $a_F$).  
We then choose $\tau_F$ as our time coordinate.  We will often write
$a_F(\tau_F)$ instead of $a_F(P(\tau_F))$;  however, one should keep in 
mind that $a_F$ and the $\tau_F(t_F)$ relation depend on the specific geodesic 
under consideration.

Now we need to define the slices of constant $\tau_F$, which requires some care.  They should not be defined by simply tracing out spatial geodesics orthogonal to the central geodesic. In fact, for a homogenous flat FLRW spacetime, parameterized by the conformal coordinates, straight lines on (3-dimensional) constant-conformal-time surfaces are {\it not} true geodesics, but are only geodesics with respect to the conformal metric, i.e. $\eta_{\mu\nu}=a^{-2} g_{\mu\nu}$. This suggests that surfaces of constant-$\tau_F$ in CFC should be spanned by space-like conformal geodesics, namely geodesics with respect to the conformal metric
\be
\tilde g_{\mu\nu}(x) \equiv a_F^{-2}(x) g_{\mu\nu}(x)\,.
\label{eq:conformal-metric}
\ee
Note that for a perturbed FLRW metric $g_{\mu\nu}=a^2 (\eta_{\mu\nu}+h_{\mu\nu})$, $\tilde g_{\mu\nu}$ in general differs from $\eta_{\mu\nu}+h_{\mu\nu}$ because the local scale factor $a_F$ is not necessarily the same as the global one $a$. Here and throughout, a tilde denotes quantities defined with respect to this conformal metric.  Let us then summarize how to locate the point $Q$ corresponding to CFC coordinates $\{\tau_F,x^i_F\}$: 

\begin{enumerate}

\item Find the point $P$ on the central geodesic corresponding to the CFC time $\tau_F$. This point thus has CFC coordinates $\{\tau_F, \bm{0}\}$.

\item Let $\tilde h(\tau_F; \alpha^i; \lambda)$ denote the family of {\it conformal geodesics} with respect to $\tilde g_{\mu\nu}$, with the affine parameter at $P$ given by $\lambda=0$ and the tangent vector at $P$ given by $\alpha^i (e_i)^\mu_P$.  That is, $\alpha^i$ are constants specifying the inital direction of the geodesic while $\lambda$ measures the geodesic distance with respect to the conformal metric (up to a constant factor). 

\item The point $Q$ is located on the conformal geodesic $\tilde h(\tau_F; a_F(P) \beta^i; \lambda)$ where $\lambda=(\delta_{ij} x^i_F x^j_F)^{1/2}$ and $\beta^i=x^i_F/\sqrt{\delta_{ij} x^i_F x^j_F}$.  This ensures that the proper distance squared from $P$ to $Q$ is $a^2_F \delta_{ij} x^i_F x^j_F$ at lowest order, which is the desired relation in the metric \refeq{CFC-requirement}.\footnote{Note that even for a curved FLRW universe, the proper distance is given by this relation to lowest order in $x_F$.}  

\end{enumerate}

This uniquely specifies the CFC coordinates, which are guaranteed to be regular in a finite region around the central geodesic (note that $a_F > 0$ is a necessary condition).   
Clearly, in order to properly specify CFC, it is not sufficient to have $a_F(\tau_F)$ given only on the geodesic.  Rather, we also need its spacetime derivatives away from the geodesic.  This is because we need to integrate 
the equation for the conformal geodesics in order to reach a point away from the geodesic, which is given by
\bea
\label{eq:conformal-geodesic-eq}
\frac{d^2 x^\mu}{d\lambda^2} + \tilde \Gamma^\mu_{\alpha\beta} \frac{dx^\alpha}{d\lambda} \frac{dx^\beta}{d\lambda} = 0\,,
\eea 
Here the Christoffel symbols for the metric $g_{\mu\nu}$ are replaced by the conformally transformed ones
\bea
\label{eq:tilGa}
\tilGa{\mu}{\alpha}{\beta} = \Gamma^\mu_{\alpha\beta} - C^\mu{}_{\alpha\beta}\,.
\eea
The tensor of shift is given by~\cite{SeanCarrollBook}
\bea
\label{eq:Ctensor}
\Cten{\mu}{\alpha}{\beta} = \delta^\mu_\alpha \,\nabla_\beta \ln a_F + \delta^\mu_\beta\, \nabla_\alpha \ln a_F - g_{\alpha\beta}\, g^{\mu\lambda} \nabla_\lambda \ln a_F\,.
\eea
In order to obtain the leading correction to the CFC metric, it is sufficient to specify the first and second derivatives of $a_F$ only.  
The first derivatives of $a_F$ are already constrained by \refeq{CFC-requirement}.  In order to obtain the CFC form, the gradient of $a_F$ along the central geodesic has to be along the time direction, i.e.
\be
\nabla_\mu \ln a_F\Big|_{x_F^i=0} = (\ln a_F)'\, a_F\,(e_0)_\mu\,,
\ee
where a prime denotes a derivative with respect to $\tau_F$.  
Note that the second derivatives $\nabla_\mu\nabla_\nu \ln a_F|_{x_F^i=0}$
are already fixed by the CFC construction, since in CFC $a_F = a_F(\tau_F)$.  
In particular, they exactly match the expression for an unperturbed FLRW
universe (\refapp{compute-conformal-connection}).

With this construction, we can now derive the explicit transformation law from some global coordinate system to CFC.  The conformal geodesic which satisfies \refeq{conformal-geodesic-eq}  has a power expansion in the curve parameter $\lambda$,
\bea
x^\mu(\lambda) = \sum^\infty_{n=0} \alpha^\mu_n \lambda^n\,.
\eea
This curve connects point $P$ with CFC coordinates $x^\mu_F(P)=\{\tau_F,\bm{0}\}$ with point $Q$ which in CFC has the coordinates $x^\mu_F(Q)=\{\tau_F,x^i_F\}$.   
Since $P$ is chosen as the spatial origin, we immediately have $\alpha^\mu_0=x^\mu(P)$.  Rescaling $\lambda$ so that it runs from $\lambda=0$ at $P$ to $\lambda=1$ at $Q$, the tangent vector at $\lambda=0$ is specified by $x_F^i$ through
\bea
\alpha^\mu_1 = \left. \frac{dx^\mu}{d\lambda} \right|_{\lambda=0} = a_F(P) (e_i)^\mu_P x^i_F\,,
\label{eq:alpha1}
\eea
Higher-order coefficients $\alpha^\mu_n$ can then be recursively computed using \refeq{conformal-geodesic-eq}.  This is shown up to third order in \refapp{coord}.

\subsection{CFC metric}
\label{sec:CFC-metric}

We now derive the explicit form that the metric attains in the CFC frame.  
For the applications we will describe below and in \cite{CFCpaper2}, we are also interested in deriving the leading correction $h^F_{\mu\nu}$ to the CFC frame metric.  

The CFC metric of course still obeys the usual tensorial transformation rule \refeq{metric-transform}.  This transformation law, evaluated at some point $Q$ not necessarily on the central geodesic, can be recast into the following form:
\bea
\label{eq:CFC-metric-conformal-transform}
\left[ (a_F(P))^{-2} g^F_{\mu\nu}\right] (x^\lambda_F) = \left[ \frac{a(P)}{a_F(P)} \frac{\partial x^\alpha}{\partial x^\mu_F} \right] \left[ \frac{a(P)}{a_F(P)} \frac{\partial x^\beta}{\partial x^\nu_F} \right] \left( \frac{a(Q)}{a(P)} \right)^2 \left[ \left(a(\tau)\right)^{-2} g_{\alpha\beta} \right] (x^\lambda),\,\,\,
\eea
so that it describes a transformation law for the conformal metric $a^{-2}g_{\mu\nu}$. The notation for the global scale factor is understood as $a(P) \equiv a(\tau(P))$ and $a(Q) \equiv a(\tau(Q))$. It is meant to be the background scale factor evaluated at the point of interest, and therefore only depends on the corresponding global time coordinate $\tau$.  This however is different for $P$ and $Q$, so that $a(P) \neq a(Q)$: different points on a constant-$\tau_F$ surface (and hence corresponding to the same $a_F(\tau_F)$ in CFC) are not simultaneous in the global coordinates, as evident from \reffig{FigureIllustration}. Therefore, while $a_F(Q)=a_F(P)$ because they refer to the same CFC time and hence the same CFC scale factor, $a(Q)$ and $a(P)$ are numerically different.

\refeq{CFC-metric-conformal-transform} can be straightforwardly evaluated
using the coordinate transformation derived in \refapp{coord}.  However, this becomes quite lengthy and is not very illuminating.  We instead present a simpler derivation through a generalization of the method adopted in Ref.~\cite{Manasse:1963zz}. The underlying idea is to project various geometric quantities into CFC, in which their components take simpler forms than in arbitrary global coordinates.

In CFC, the temporal coordinate vector is $(\tilde e_0) \equiv \partial/\partial x^0_F = a_F(P) (e_0)_P$, and the spatial coordinate vectors are $(\tilde e_i) \equiv \partial/\partial x^i_F = a_F(P) (e_i)_P$. Hence the CFC metric right on the central geodesic must be conformally Minkowskian,
\bea
\label{eq:eq-CFC-zeroth}
\left.g^F_{\mu\nu}\right|_P = \left[ \left(\tilde e_\mu\right)^\alpha \left(\tilde e_\nu\right)^\beta g_{\alpha\beta} \right]_P = a^2_F(P) \left[ \left(e_\mu\right)^\alpha \left( e_\nu\right)^\beta g_{\alpha\beta} \right]_P = a^2_F(P)\,\eta_{\mu\nu},
\eea
as the tetrad vectors are orthonormal.   Further, the spatial CFC coordinate lines are geodesics of the conformal metric, and are thus simply parametrized through 
\bea
\label{eq:eq-CFC-spatial-geo-par}
x^0_F = \tau_F = {\rm const.},\qquad x^i_F = \beta^i \lambda\,,
\eea 
where $\beta^i$ are constants and $\lambda$ is the affine parameter.  
Since the tangent vector is $(0,\beta^i)$, the conformal geodesic equation \refeq{conformal-geodesic-eq} reduces in CFC to
\bea
\left. (\tilde \Gamma^F )^\mu_{ij} \right|_P \,\beta^i \beta^j = 0,
\eea
which implies $(\tilde \Gamma^F)^\mu_{ij} |_P=0$ because $\beta^i$ is arbitrary. 
One can further show easily (\refapp{CFC-metric-CFC}) that following \refeq{eq-CFC-spatial-geo-par}, all Christoffel connection coefficients $(\tilde \Gamma^F)^\mu_{\alpha\beta}$, computed with respect to $\tilde g_{\mu\nu}$ [\refeq{conformal-metric}] and {\it in CFC}, vanish on the central geodesic, i.e. $(\tilde \Gamma^F)^\mu_{\alpha\beta} |_P = 0$. It follows that all first-order derivatives of the conformal metric $\tilde g_{\mu\nu}$ in CFC vanish on the central geodesic,
\bea
\label{eq:eq-CFC-first-deriv}
\left. \left(\partial_\alpha\, \tilde g_{\mu\nu}\right)^F \right|_P = 0\,.
\eea 
This proves the absence of $\mathcal{O}[x^i_F]$ terms in the CFC metric.

Furthermore, the $\mathcal{O}[(x^i_F)^2]$ terms in the CFC metric are found to be related to the {\it conformal} Riemann curvature tensor $\tilde R$, i.e. the Riemann tensor constructed for the metric \refeq{conformal-metric}, and evaluated on the central geodesic. Given \refeq{eq-CFC-zeroth} and \refeq{eq-CFC-first-deriv}, the entire derivation in Ref.~\cite{Manasse:1963zz}, which is for an analogous relation between the FNC quadratic metric terms and the ordinary Riemann curvature tensor, can be borrowed over, once conformally-related quantities replace the ordinary ones everywhere (e.g. metric, geodesic equation, geodesic deviation equation, Christoffel symbols, Riemann curvature tensors, and so on). The end result for the quadratic corrections reads
\bea
\label{eq:gF-00}
g^F_{00}(x_F) & = & a^2_F(\tau_F) \left[ -1 - \left.\tilde R^F_{0k0l} \right|_P x^k_F x^l_F \right], \\
\label{eq:gF-0i}
g^F_{0i}(x_F) & = & a^2_F(\tau_F) \left[ - \frac23 \left.\tilde R^F_{0kil} \right|_P x^k_F x^l_F \right], \\
\label{eq:gF-ij}
g^F_{ij}(x_F) & = & a^2_F(\tau_F) \left[ \delta_{ij} - \frac13 \left.\tilde R^F_{ikjl} \right|_P x^k_F x^l_F \right].
\eea
Here $\tilde R^F$ is the Riemann curvature tensor constructed with respect to $\tilde g_{\mu\nu}$ and transformed to the CFC frame. In terms of some global coordinates,
\bea
\tilde R^F_{\alpha\beta\gamma\delta} & = & (\tilde e_\alpha)^\mu (\tilde e_\beta)^\nu (\tilde e_\gamma)^\rho (\tilde e_\delta)^\sigma \tilde R_{\mu\nu\rho\sigma},
\eea
where $\tilde R_{\mu\nu\rho\sigma}$ is the Riemann tensor of the conformal metric with its components computed in the global coordinates.  
Note that the indices of the conformal Riemann tensor $\tilde R$ are
always to be raised and lowered with the \emph{conformal} metric $\tilde g_{\mu\nu}$.

Extending the analysis to higher orders, as detailed in \refapp{CFC-metric-CFC}, the $\mathcal{O}[(x^i_F)^3]$ corrections to the CFC conformal metric scale as
\be
\left(\partial_i \tilde R^F_{\mu j \nu k} \right)_P x_F^i x_F^j x_F^k\,,
\ee
i.e. they are suppressed by \emph{the spatial derivative of $\tilde R^F$ multiplied by $x_F$}.  Further, the coefficient of the $\mathcal{O}[(x^i_F)^4]$
terms in general contains $\partial_i\partial_j \tilde R^F$ as well as
$(\tilde R^F)^2$.  
This illustrates that we are expanding in the spatial variation of the
tidal force $\tilde R^F$ induced by the long mode, as well as its amplitude.

\subsection{Choosing the CFC scale factor}
\label{sec:choice-aF}

So far we have not written down any equation that determines the spacetime scalar $a_F(x)$.  Note however that since in CFC $a_F$ can only depend on $x_F^i$ at second or higher order, and since we fixed the second order term by matching to FLRW solution, we have effectively reduced this freedom to a function $a_F(\tau_F)$ of time only.  
Thus, another way of saying this is that the aforementioned construction is invariant under the most general reparameterization of the conformal time,
\bea
\tau_F = \tau_F(\tau'_F), \qquad a_F(\tau_F) = a'_F(\tau'_F) \frac{d \tau'_F}{d \tau_F},
\eea
under which $a_F(\tau_F) d\tau_F = dt_F$ is still the proper time interval along the central geodesic. The new $x^{i\prime}_F$'s then have to be re-defined following our procedure using conformal geodesics, which in general leads to a different slicing of simultaneity, and different CFC metric perturbations \refeqs{gF-00}{gF-ij}.  Note in particular that we can simply set $a_F = 1$, in which case we recover the standard FNC frame described in \refsec{FNC}.  

Let us investigate this issue a bit further.  Consider a congruence of geodesics in the vicinity of the chosen central geodesic, and let $U^\mu= dx^\mu/dt_F$ denote the tangent vector to this congruence.  Note that $(e_0)^\mu = U^\mu$ defines the time component of our tetrad.  
$U^\mu$ satisfies the geodesic equation $U^\alpha\nabla_\alpha U^\mu=0$, 
and hence $\nabla_\mu \left(U^\alpha \nabla_\alpha U^\mu \right)=0$.  Straightforward manipulation allows us to derive 
\ba
U^\alpha  \nabla_\alpha ( \nabla_\mu U^\mu ) =\:& 
- (\nabla_\mu U^\alpha) \,(\nabla_\alpha U^\mu) 
-  R^\mu_{\  \beta\mu\alpha} U^\beta U^\alpha \,.
\label{eq:step1}
\ea
Because $U^\mu$ is geodesic, the velocity shear tensor $\nabla_\mu U^\alpha$ is in the ``velocity-orthogonal'' 3-dimensional subspace, whose projector is given by
\be
P^\mu_{\  \nu} = \delta^\mu_{\  \nu} + U^\mu U_\nu\,,
\label{eq:proj}
\ee
noting that we use the mostly positive metric convention;  explicitly, the geodesic equation implies that $P^\mu_{\  \nu} P^\alpha_{\  \beta} \nabla^\nu U^\beta = \nabla^\mu U^\alpha$.  
We now decompose the velocity shear tensor in the usual way as
\ba
\nabla_\mu U^\alpha =\:& \frac13 \vartheta P_\mu^{\  \alpha}
+ \sigma_\mu^{\  \alpha} + \omega_\mu^{\  \alpha}\,,
\ea
where $\vartheta = P^\mu_{\  \alpha} \nabla_\mu U^\alpha = \nabla_\mu U^\mu$ is the
velocity divergence, 
$\sigma$ is the trace-free symmetric velocity shear, while
$\omega$ is the antisymmetric part, i.e. vorticity.  That is,\footnote{Throughout we are always raising and lowering indices with $g_{\mu\nu}$, not $P_{\mu\nu}$.} $\sigma_{\alpha\beta} = \nabla_{(\mu} U_{\alpha)} -  \vartheta P_{\alpha\beta}/3$.  
\refeq{step1} then becomes, using $U^\alpha\nabla_\alpha = d/dt_F$ and $P_{\alpha\beta} P^{\alpha\beta} = 3$, and using the antisymmetry of $\omega_{\alpha\beta}$,
\ba
\frac{d}{dt_F} \vartheta =\:& 
- \frac13 \vartheta^2 - \sigma_{\alpha\beta} \sigma^{\alpha\beta}
+ \omega_{\alpha\beta} \omega^{\alpha\beta}
-  R^\mu_{\  \beta\mu\alpha} U^\beta U^\alpha \,.
\label{eq:raych}
\ea
This is the well-known Raychaudhuri equation, where typically the contraction of the Riemann tensor is replaced with the Ricci tensor.  We will not do this because we prefer results in terms of the Riemann tensor.  Note that \refeq{raych} is a purely geometric result; we have not made use of Einstein's equations.  Working in CFC, we have $U^\mu = a_F^{-1}(1,0,0,0)$, so that \refeq{raych} becomes
\ba
\frac{d}{dt_F} \vartheta =\:& 
- \frac13 \vartheta^2 - \sigma_{ij} \sigma^{ij}
+ \omega_{ij} \omega^{ij}
-  a_F^{-2} (R^F)^\mu_{\  0\mu0} \,.
\ea
Using the transformation law of the Riemann tensor under a conformal rescaling of the metric, $g_{\mu\nu} \to \tilde g_{\mu\nu} = a_F^{-2} g_{\mu\nu}$, and some algebra we obtain
\ba
a_F^{-2} (R^F)^\mu_{\  0\mu 0} 
= a_F^{-2} (\tilde R^F)^\mu_{\  0\mu 0} - 3 (\dot H_F + H_F^2) \,.
\ea
Here we have used $()' = a_F (\dot{\  })$ when acting on a simple function of time along the geodesic.  We can also use the fact that by way of the CFC construction,
$(\tilde R^F)^0_{\  0\mu 0} = 0$ (this is because the tetrad is parallel-transported; see \refsec{CFC-lagrange-eulerian}).   
Finally, we obtain the Raychaudhuri equation in CFC:
\ba
\frac{d}{dt_F} \vartheta + \frac13 \vartheta^2 =\:& 3 (\dot H_F + H_F^2)
 - \sigma_{ij} \sigma^{ij}
+ \omega_{ij} \omega^{ij}
-  a_F^{-2} (\tilde R^F)^j_{\  0 j 0} \,.
\label{eq:raychF}
\ea
Again, we have made no assumption about the spacetime or about $a_F$ so far.  We only used the requirement that $U^\mu$ is tangent to a congruence of geodesics.  In other words, for any spacetime we can always choose coordinates where the Raychaudhuri equation takes this form.

Let us drop $\sigma$ and $\omega$ for the moment (in the case of perturbed FLRW discussed in \refsec{perturbed-FLRW}, they are at least linear in perturbations, so that their contribution to \refeq{raychF} is at least second order and hence will be dropped in our linear treatement there).  
We then have 
\ba
\frac{d}{dt_F} \vartheta + \frac13 \vartheta^2 =\:& 3 (\dot H_F + H_F^2)
-  a_F^{-2} (\tilde R^F)^j_{\  0j 0} \,.
\label{eq:raychF2}
\ea
Recall that once we have picked our central geodesic, this equation holds at all points along that geodesic, and all quantities are just functions of time.  
We can now use our freedom to choose $a_F(t_F)$ as we desire.  For example,
choosing $a_F = 1$ reduces $\tilde R^F$ to $R^F$, which then contains the entire source of local velocity divergence.  Our goal is to absorb the leading contributions from the Hubble expansion in $a_F$ however, and make $\tilde R^F$ ``as small as possible''.  Guided by this principle, we will define $a_F$ through
\be
H_F = \frac{d\ln a_F}{dt_F} := \frac13 \vartheta
\label{eq:vartheta-def}
\ee
throughout this paper and the companion paper \cite{CFCpaper2}.  
This is a convenient choice, first of all because it reduces to the true scale factor in the case of an unperturbed FLRW spacetime, in which case $\tilde R^F$ then vanishes entirely.  Second, $\vartheta$ is then an \emph{observable along the geodesic}\footnote{Note that $\vartheta$ is not strictly a \emph{local} observable, since it requires a time integration.  The true local observable is $d\vartheta/dt_F$.  Nevertheless, it can be reconstructed by a local observer if we follow his geodesic through time.  Our choice neglects any initial peculiar velocities of the geodesic congruence, which decay $\propto a^{-1}$ due to the Hubble expansion.} and thus $H_F$ is as well,  
and third, \refeq{raychF2} implies
\be
(\tilde R^F)^j_{\  0j 0} = 0\,.
\label{eq:RFjj}
\ee
That is, the second spatial derivatives of the CFC metric correction [\refeq{gF-00}] are \emph{trace-free} and correspond to a pure tidal field.  \refeq{RFjj} of course only follows from \refeq{vartheta-def} in the absence of velocity shear and vorticity.  More generally, our condition \refeq{vartheta-def} implies
\ba
a_F^{-2} (\tilde R^F)^j_{\  0j 0}  =\:& 
- \sigma_{ij} \sigma^{ij}
+ \omega_{ij} \omega^{ij}\,.
\label{eq:raychFgauged}
\ea
So, one way of interpreting our choice of $a_F$ is that it brings the Raychaudhuri equation into the simple form \refeq{raychFgauged}, which algebraically relates the trace part of $h^F_{00}$ to the velocity shear and vorticity.  

Note that \refeq{vartheta-def} only determines $a_F$ up to a multiplicative constant, which can be trivially absorbed in a constant rescaling of the spatial coordinates $x_F^i$.

\subsection{Residual gauge freedom at $\mathcal{O}[(x^i_F)^2]$}
\label{sec:residual-freedom-xF2}

Even when combined with a prescription to fix $a_F(\tau_F)$, the CFC construction does \textit{not} univocally determine the CFC metric perturbations $h^F_{\mu\nu}$. The situation is in fact analogous to the case of FNC (see App.~D of~\cite{Pajer:2013ana}). The metric components $h^F_{00}$ and $h^F_{0i}$ are completely fixed at $\mathcal{O}[(x^i_F)^2]$ by the requirement of conformal flatness at $\mathcal{O}[x^i_F]$. However, a residual gauge freedom exists for the spatial components $h^F_{ij}$. Consider the following (time-dependent) reparameterization of $x^i_F$:
\bea
x^i_F & \longrightarrow & x^i_F + \frac16 A^i{}_{jkl}\left(\tau_F\right)\, x^j_F x^k_F x^l_F + \mathcal{O}[(x^i_F)^4],
\eea
where the coefficient tensor $A^i{}_{jkl}\left(\tau_F\right)$ is an arbitrary function of the CFC time and is fully symmetric with respect to its last three indices. This change of coordinates leads to the following change in $h^F_{ij}$:
\bea
\label{eq:residual-gauge-hFij}
h^F_{ij} & \longrightarrow & h^F_{ij} + A_{(i,j)kl}\, x^k_F x^l_F+ \mathcal{O}[(x^i_F)^3].
\eea 
This residual gauge can be used to bring $h^F_{ij}$ into a desirable shape.  
As an example, one can think of the different ways to
parametrize spatial hypersurfaces of constant curvature, e.g.
conformally flat and stereographic projections. Note that $A^i_{\  jkl}$ has to drop out from the final result for any given physical observable. We have verified that this is the case for the application we will consider in \refsec{long-tensor}.

\section{Perturbed FLRW spacetime}
\label{sec:perturbed-FLRW}

The CFC construction described in the previous section is completely
general, i.e. it can be applied to any spacetime.  However, the most obvious
application is the case of a perturbed FLRW spacetime, given in some global
coordinates by
\bea
\label{eq:perturbed-FLRW}
ds^2 = a^2(\tau) \left[ \eta_{\mu\nu} + h_{\mu\nu} \right] dx^\mu dx^\nu\,.
\eea
For full generality we do not make a specific gauge choice for the metric perturbation $h_{\mu\nu}$.  However, in the following we assume that $h_{\mu\nu}$ is small, so that it suffices to work at linear order in $h$.  The perturbative expansion in $h_{\mu\nu}$ should not be confused with the power expansion in $x^i_F$.  The former expansion is valid as long as $|h_{\mu\nu}| \ll 1$, and adopted here merely for calculational simplicity, whereas the latter expansion is valid if $|x^i_F|$ is smaller than the typical scale of variation of $h_{\mu\nu}$, and is the true inherent regime of validity of the CFC frame.   We leave a generalization to nonlinear order in $h_{\mu\nu}$ to future work.  In realistic cosmological settings, metric perturbations are present on all scales, so that a coarse-graining needs to be performed in order for the CFC to be valid over a finite region.  We return to this issue in \refsec{EE}.


Consider a free-falling observer traveling along the time-like central geodesic. Her 4-velocity can be parameterised as $U^\mu = a^{-1} \left(1+h_{00}/2,V_o^i\right)$, where the 3-velocity $V^i_o$ is considered as first-order perturbation. The corresponding tetrad is\footnote{The expressions here for the tetrads, as well as other quantities $V^i_o,\,\omega^i,\cdots$, are not restricted to the central geodesic, as one can think of imaginary free-falling observers following through every time-like geodesic in the Universe. Therefore, partial derivatives of them are meaningful.}
\bea
(e_0)^\mu & = & a^{-1} \left( 1 + \frac12 h_{00}, V^i_o \right), \\
(e_i)^\mu & = & a^{-1} \left( V_{o,j} + h_{0j}, \delta^i{}_j - \frac12 h^i{}_j + \frac12 \varepsilon_j{}^{ik} \omega_k \right),
\eea 
where $\varepsilon_{ijk}$ is the three-dimensional Levi-Civita tensor.  Here and in the following, latin indices $i,j,k,...$ are raised and lowered with $\delta_{ij}$ following standard practice in cosmology.  In the presence of vector metric perturbations, it is necessary to introduce the rotation $\omega^i$.  Since the tetrad is parallel-transported along the central geodesic, $V^i_o$ and $\omega^i$ obey the following equations,
\bea
\label{eq:CFC-obs-eom}
V^{i\prime}_o + \coH V^i_o & = & \frac12 \partial^i h_{00} - h^i_{\  0}{}' - \coH h^i{}_0, \\
\omega^{k\prime} & = & -\frac12 \varepsilon^{kij} \left( \partial_i h_{0j} - \partial_j h_{0i} \right).
\eea  
A prime denotes derivative with respect to the conformal time.


We now determine the CFC scale factor $a_F$ through the definition \refeq{vartheta-def}.  
In the perturbed spacetime \refeq{perturbed-FLRW} $\vartheta$ is given by
\bea
\label{eq:congruence-central-geodesic}
\vartheta = \frac 1a \left[ 3 \coH + \frac32 \coH h_{00} + \frac12 h' + \partial \cdot V_o \right],
\eea
where $h=\delta^{ij} h_{ij}$. It is understood that the right hand side is computed in global coordinates on the central geodesic. 
The local Hubble parameter is then given by
\bea
\label{eq:local-Hubble-2}
\frac{\coH_F(\tau_F)}{a_F(\tau_F)} = \frac{1}{a(\tau)} \left[ \coH(\tau) + \frac12 \coH(\tau) h_{00} + \frac16 h' + \frac13 \partial \cdot V_o \right].
\eea
Moreover, the ``cosmic acceleration'' is
\bea
\label{eq:eq-CFC-acceleration}
\frac{1}{a^2_F(\tau_F)} \frac{d\coH_F(\tau_F)}{d\tau_F} = \frac{1}{a^2_F(\tau_F)} \frac{d}{d\tau_F} \left( \frac{1}{a_F(\tau_F)} \frac{d a_F(\tau_F)}{d\tau_F} \right) = \frac13 \frac{\coH_F(\tau_F)}{a_F(\tau_F)} \,\vartheta + \frac13 (e_0)^\mu_P \,\partial_\mu \vartheta\,.
\eea

Even after the condition \refeq{vartheta-def}, there is a residual freedom in $a_F$, corresponding to a constant rescaling $a_F \to c\,a_F$ (equivalent to an integration constant when integrating the Hubble rate). This (trivial) residual freedom can be most conveniently removed by requiring $a_F$ to asymptote to $a$ when compared at fixed CFC proper time $t_F$,
\bea
\frac{a_F(\tau_F(t_F))}{a(\tau(t=t_F))} \rightarrow 1, \qquad {\rm as}\,\,\, t_F \rightarrow 0\,.
\eea
For a matter-dominated universe at early times, this implies
\bea
\frac{a_F(\tau_F)}{a(\tau)} \rightarrow 1 - \frac13\, h_{00,{\rm ini}},\qquad {\rm as}\,\,\, \tau_F \rightarrow 0\,.
\eea
where $h_{00,\rm ini} = h_{00}(\tau_{\rm ini})$ is the metric perturbation evaluated on the geodesic at early times on superhorizon scale. \refeq{local-Hubble-2} yields
\bea
\frac{d}{d\tau} \frac{a_F(\tau_F)}{a(\tau)} = - \frac{a_F \coH}{a} + \frac{a_F \coH_F}{a} \frac{d\tau_F}{d\tau} = \frac{a_F}{a} \left[ \frac16 h' + \frac13 \partial \cdot V_o \right]\,,
\eea
and a direct integration gives
\bea
\frac{a_F(\tau_F)}{a(\tau)} = 1 - \frac13 H\, h_{00,{\rm ini}} + \int^\tau_{\tau_{\rm ini}} d\tau \left( \frac16 h' + \frac13 \partial \cdot V_o \right)\,.
\label{eq:aF}
\eea
This defines the physical, locally inferred CFC scale factor for a general
linearly perturbed FLRW metric.

Our construction of the CFC is almost identical to that of \cite{Pajer:2013ana}, with one important difference in the definition of $a_F$.  Here, we 
have constructed $a_F$ as (essentially) a local observable, while 
in \cite{Pajer:2013ana} $a_F$ was chosen to be equal to the background scale factor $a$, i.e.
$a_F(t_F) = a(t=t_F)$.  This choice however does not correspond to an
observable scale factor and Hubble rate $\coH_F$ (since $a_F$ differs for
different gauge choices in global coordinates).  In this case, the CFC metric
corrections $\propto \tilde R^F_{\mu k \nu l}$ in \refeqs{gF-00}{gF-ij} are not guaranteed to be observable
by themselves.  On the other hand, by ensuring that $a_F$ is observable,
these corrections are observable, which is a crucial advantage.  Nevertheless,
the choice made in \cite{Pajer:2013ana}  was sufficient for
their conclusions, since their goal was to show the absence of long-wavelength
corrections to the CFC frame at lowest order in derivatives (not suppressed
by $k_L$).  On the other hand, for long-wavelength tensor perturbations, 
which is the subject of \cite{Schmidt:2013gwa} and \refsec{long-tensor},
$a_F$ is equal to $a$ at linear order in the tensor amplitude.

\section{Coarse-graining and Einstein equations in the CFC frame}
\label{sec:EE}

Having discussed the definition of the CFC frame and metric for a perturbed
FLRW spacetime, in this section we show how this construction removes gauge
artifacts and simplifies the calculation of nonlinear long-short mode coupling in the regime of practical interest in cosmology.

The corrections $h^F_{\mu\nu}$ to FLRW in the CFC metric are composed of 
second derivatives of the global metric perturbations $a^2 h_{\mu\nu}$ multiplied by $(x_F^i)^2$.  Specifically, we have terms of order 
\be
h^F_{\mu\nu} \sim \Big\{ h_{\mu\nu}'',\; \coH h_{\mu\nu}',\; \partial_k h_{\mu\nu}',\; \partial_k \partial_l h_{\mu\nu} \Big\} x_F^i x_F^j\,.
\ee
The corrections $h^F_{\mu\nu}$ are locally observable, as the CFC construction makes explicit (note that a physical definition of a locally measurable scale factor is crucial for this).  Thus, the equivalence principle guarantees that for a single Fourier
mode $k_L$ of an adiabatic perturbation, $h^F_{\mu\nu}$ scales as $k_L^2$ in the limit $k_L \to 0$.  Note however that individual contributions to $h^F_{\mu\nu}$ in some global coordinate system are not in general observable by themselves; terms of the type $\coH h_{\mu\nu}'$ listed above are an example.  
Higher order corrections to the CFC metric which we have dropped are correspondingly
suppressed by higher derivatives of $h_{\mu\nu}$ multiplying higher powers of
$x_F^i$ (\refsec{CFC-metric}).  This means that the CFC metric is only valid over comoving spatial scales much smaller than $\lambda_L = k_L^{-1}$.  

In our universe, metric perturbations exist down to very small scales. For the CFC metric to be valid over a finite region, we need to introduce a coarse-graining of the metric on some comoving spatial scale $L = \L^{-1}$.  The CFC metric
is then constructed with respect to the coarse-grained metric perturbation
$h^\L$, which in Fourier space has contributions only for wavenumbers
$k \lesssim \L$.  In this section,
we describe the coarse-graining and derive the structure of Einstein equations in the CFC frame.  We work throughout to linear order in the
coarse-grained metric perturbation, but allow the small-scale perturbations
to be nonlinear.

\subsection{Coarse graining}

Consider the Einstein equations written in some global coordinate system:
\be
G_{\mu\nu}[g] = 8\pi G \, T_{\mu\nu}\,.
\label{eq:EE}
\ee
We assume everywhere that a cosmological constant is defined into $G_{\mu\nu}$ (or equivalently $T_{\mu\nu}$) for simplicity.  
Specializing to the case of a perturbed FLRW metric [\refeq{perturbed-FLRW}],
we subtract the homogeneous background solution, to obtain the Einstein
equation for the metric perturbation $h_{\mu\nu}$:
\be
G^{a(\tau)}_{\mu\nu}[h] = 8\pi G \, T_{\mu\nu}\,,
\label{eq:EE0}
\ee
where $G^{a(\tau)}[h]$ is the Einstein tensor for the perturbation to the
FLRW metric, which depends on $a(\tau)$.  $T_{\mu\nu}$ now only contains the
perturbation to the stress energy tensor.  Since we no longer need the
homogeneous background stress energy tensor, we have kept the same symbol $T_{\mu\nu}$
for simplicity.  

We now coarse-grain the stress-energy tensor and the metric on a spatial scale
$\L^{-1}$.  This means we have to introduce a slicing of spacetime.  The choice of slicing influences the results obtained in the coarse-grained CFC calculation, but it does not affect the structure of the equations, which is our main focus here.  For simplicity, we will assume that the coarse-graining is performed on a constant-coordinate-time slice.  Hence, we write
\be
h_{\mu\nu} = h_{\mu\nu}^\L + h_{\mu\nu}^s; \quad
T_{\mu\nu} = T_{\mu\nu}^\L + T_{\mu\nu}^s\,, 
\label{eq:sLambda}
\ee
where the ``s'' components are defined as the difference between the full
quantity and its coarse-grained version, i.e. through \refeq{sLambda}.  
Let us further separate the
Einstein tensor into linear and nonlinear pieces in the metric perturbation, 
$G^{a(\tau)}[h] = G^{(\rm lin)}[h] + G^{(\rm nl)}[h]$,
where $G^{(\rm lin)}$ is the part of the Einstein
tensor that commutes with the coarse graining.

From coarse-graining \refeq{EE0}, we obtain
\be
G^{(\rm lin)}_{\mu\nu}[h^\L] + G^{(\rm nl) \L}_{\mu\nu}[h] = 8\pi G \, T^\L_{\mu\nu}\,,
\ee
which can be turned into an implicit equation for $h^\L$,
\be
G^{(\rm lin)}_{\mu\nu}[h^\L] + G^{(\rm nl)}_{\mu\nu}[h^\L] = 8\pi G \, T^\L_{\mu\nu}
+ \left\{G^{(\rm nl)}_{\mu\nu}[h^\L] - G^{(\rm nl) \L}_{\mu\nu}[h]\right\}
\,.
\label{eq:EEL}
\ee
The term in curly brackets on the r.h.s. is an effective source in the
Einstein equation for $h^\L$.  It encapsulates the backreaction of small
scales, which are contained in the full $h= h^\L + h^s$, on large scales.  In the 
following, we will neglect this term, since we only work to linear order in 
$h^\L$, and we are not interested in the backreaction effect of small scales on large
scales (which in general is a difficult problem).  

Let us now subtract \refeq{EEL} from \refeq{EE0},
dropping the additional source, to obtain an equation for $h^s$:
\ba
G_{\mu\nu}^{(\rm lin)}[h^s] + G_{\mu\nu}^{(\rm nl)}[h] - G_{\mu\nu}^{(\rm nl)}[h^\L] =\:& 8\pi G\, T^s_{\mu\nu}\,, 
%
\label{eq:EEs}
\ea
We are interested in the mode-coupling between long-wavelength modes $h^\L$
and short scales $h^s$.  Suppose we have a solution $h$ to the full
Einstein equations \refeq{EE0}.  The small-scale part of the solution $h^s$
can be expanded as power series in $h^\L$:
\be
h^s = h^{s0} + h^{s\L} + h^{s\L^2} + \cdots\,,
\label{eq:hsseries}
\ee
where $h^{s\L^n}$ is proportional to $(h^\L)^n$.  Thus, $h^{s0}$ is the small-scale
metric perturbation that remains in the limit of vanishing long-wavelength
metric perturbations $h^\L \to 0$.  
In keeping with the
linear treatment of $h^\L$, we will truncate the series at $h^{s\L}$ in the
following. 
Note that $T^s$ is defined through \refeq{sLambda}.  The stress-energy is
of course a nonlinear function of the fluid variables and the metric.  
For example, a component such as $p g_{\mu\nu}$ in $T^s_{\mu\nu}$ is given by
\be
(p g_{\mu\nu})^s = p g_{\mu\nu} - (p g_{\mu\nu})^\L\,.
\label{eq:TsLex}
\ee
Similar relations hold for say $\rho u_\mu u_\nu$.  Thus, $T^s_{\mu\nu}$ is to be expanded in a power series of
the type \refeq{hsseries} up to the same order as $h^s$.   At zeroth order,
the Einstein equation becomes, not surprisingly,
\be
G_{\mu\nu}^{(\rm lin)}[h^{s0}] + G_{\mu\nu}^{(\rm nl)}[h^{s0}] = 8\pi G\, 
T^{s0}_{\mu\nu} \,,
\label{eq:EEs0}
\ee
while at linear order in $h^\L$ we obtain the equation for $h^{s \L}$: 
\ba
G_{\mu\nu}^{(\rm lin)}[h^{s \L}] + G^{(\rm nl)\prime}_{\mu\nu}[h^{s0}, h^\L + h^{s\L}] 
=\:& 8\pi G\, T^{s\L}_{\mu\nu}\,, \label{eq:EEsL}
\ea
where the ``response'' operator is defined by
\ba
G^{(\rm nl)\prime}[h, \tilde h] =\:& \frac{\partial G^{(\rm nl)}[h]}{\partial h_{\mu\nu}} \tilde h_{\mu\nu} +
\frac{\partial G^{(\rm nl)}[h]}{\partial(\partial_\alpha h_{\mu\nu})} \partial_\alpha \tilde h_{\mu\nu} +
\frac{\partial G^{(\rm nl)}[h]}{\partial(\partial_\alpha \partial_\beta h_{\mu\nu})} \partial_\alpha \partial_\beta \tilde h_{\mu\nu}\,.
\label{eq:Gnlprime}
\ea
This operator is obviously linear in its second argument and corresponds to
the linear term in the Taylor series of $G^{({\rm nl})}$ around $h$ (we have dropped the tensor indices for clarity).  
As desired, \refeq{EEsL} is linear in $h^\L$ and $h^{s\L}$, but in general 
it is nonlinear in $h^{s0}$.  This is the equation one is solving when
computing the impact of a linear large-scale perturbation $h^\L$ on small-scale
perturbations in General Relativity.

\subsection{Transforming to CFC frame}

\refeq{EEsL} was derived in some global coordinate system.  Our goal now
is to obtain the analogous equation in the CFC frame.  Specifically, 
we construct the CFC frame for the \emph{coarse-grained} metric.  
Of course, the full
Einstein equations \refeq{EE0} transform covariantly.  However, we
have performed a coarse-graining and ``$\L-s$'' split which breaks
covariance. Let us thus look at the transformation more carefully.  
The transformation to CFC
\be
x \to x_F(x)\,,
\ee
is by assumption a ``long-wavelength'' one, since it is purely determined by the 
coarse-grained metric.  That is, we can write
\be
\frac{\partial x^\mu}{\partial x_F^\alpha} = \delta^\mu_{\  \alpha} + A^\mu_{\  \alpha}[h^\L]\,,
\ee
where $A$ is linear in $h^\L$.  Any symmetric tensor transforms as
\ba
G_{\mu\nu}^F =\:& \frac{\partial x^\alpha}{\partial x_F^\mu} \frac{\partial x^\beta}{\partial x_F^\nu} G_{\alpha\beta}
= G_{\mu\nu} + 2 A^{\alpha}_{\  (\mu} G_{\alpha \nu)} + \O(A^2)\,,
\label{eq:Gtrans}
\ea
where $G^F$ denotes the tensor in the CFC frame,
and we can neglect the $\O(A^2)$ part in keeping with our linear treatment
in $h^\L$.  In particular, applying this to the long-wavelength metric,
we have by construction
\be
\frac{\partial x^\alpha}{\partial x_F^\mu} \frac{\partial x^\beta}{\partial x_F^\nu} a^2 \left[ \eta_{\alpha\beta} + h_{\alpha\beta}^\L \right]
= a^2 \left[\eta_{\mu\nu} + h_{\mu\nu}^\L + 2  A_{\mu\nu}\right]
= a_F^2 \left[\eta_{\mu\nu} + h_{\mu\nu}^{\L,F} \right]\,,
\ee
where $h_{\mu\nu}^{\L,F} \propto x_F^i x_F^j$ in the vicinity of $\vx_F=0$.  
It is then easy to see that each term in \refeq{EEs} and \refeq{EEsL} transforms
in the same way, so that the equation for $h^{s\L}$ in CFC reads
\ba
G^{(\rm lin) F}[h^{s\L}_F] +  G^{(\rm nl)F \prime}[h^{s0},h^{s\L}_F]
+ G^{(\rm nl)F \prime}[h^{s0},h^\L_F ]  =\:& 8\pi G\, T^{s\L,F} \,,
\label{eq:EEsLC}
\ea
where we have suppressed the tensor indices for simplicity.  Here,
the tensors $G^{(\rm lin) F}$ and $G^{(\rm nl)F\prime}$ are contructed
for the CFC metric, in particular they involve the local scale factor
$a_F$ in the background.  Thus, the
Einstein equation for $h_F^{s\L}$ in the CFC
frame  has the same structure as the one written in global coordinates
\refeq{EEsL}.

\subsection{CFC Einstein equations}

Let us now investigate \refeq{EEsLC}, focusing on the l.h.s. for the
time being.  In the following, we will only deal with 
the CFC metric. Therefore, for the sake of clarity, we will omit the ``F'' label (referring to CFC) everywhere, except in $h^\L_F$, which differs
crucially from $h^\L$ in global coordinates.  
Suppose further that we are aiming to solve
the Einstein equation \emph{on the central geodesic} $\vx_F=0$.  We will
return to this point below, but note that we are free to construct the CFC around any timelike geodesic.  
We will study two regimes in which the analysis simplifies in different
ways.  First, we consider the case of a second order treatment in 
cosmological perturbations, corresponding to the leading order contribution
to the mode-coupling term $h^{s\L}$.  Second, we will consider the general
nonlinear case in small-scale perturbations, but restrict to the case
where the small-scale fluctuations are far inside the Hubble horizon.  
Since in the standard cosmology perturbations are small outside the horizon and become nonlinear only well inside it, this completely encompasses the regime of structure formation.  

\subsubsection{Second order in perturbations}

Restricting to second order in perturbations, the second term on 
the l.h.s. of \refeq{EEsLC} is higher order, as $h^{s\L}$ is already second 
order.  Let us then consider the third term on the l.h.s..  Since
we need at least two spatial derivatives on $h^\L_F$ to obtain a nonzero 
contribution on the geodesic, only a small specific subset of
terms contribute to $G^{(\rm nl)\prime}[h^{s0},h^\L_F]$ in \refeq{Gnlprime}.  
We can write \refeq{EEsLC} as
\ba
\mbox{CFC, 2nd order:}\quad G_{\mu\nu}^{(\rm lin)}[h^{s \L}] 
+ \O( h^{s0} \partial_i \partial_j h^\L_F )_{\mu\nu} =\:& 8\pi G
T^{s\L}_{\mu\nu} \,.
\label{eq:EEsLC2}
\ea
Note that since the Einstein tensor is linear in second derivatives of 
the metric, $h^{s0}$ has to enter without derivatives here.  
In order to solve the Einstein equation at second order in CFC for $\vx_F=0$,
we thus only have to work out a very specific type of terms, i.e. 
those with two perturbations and two spatial
derivatives acting on one of them.  

One caveat to this statement is that this assumes we do not act
with spatial derivative operators on the Einstein equations.  
In practice, we might want to do that to extract the 
longitudinal part of the $ij$ component of the Einstein equations.  
In that case, we need to include
the contributions of $\O[x^i]$ and $\O[(x^i)^2]$ to the CFC Einstein
equations, while any higher powers of $x^i$ are consistently dropped
in the CFC expansion (\refsec{CFC-metric}). This diminishes the calculational simplicity
of the CFC Einstein equations when $k_S$ is superhorizon.  

\subsubsection{Subhorizon limit}
\label{sec:EESH}

Let us now consider the subhorizon limit for the small-scale perturbations $h^s$, i.e. $k_S \gg \coH$.  As always,   
we allow the long-wavelength perturbation $h^\L$ to be arbitrarily long.  In the subhorizon limit, we approximate
\be
\partial_i\partial_j h^s \sim 1,\quad
\partial_i h^s \sim \eps_s,\quad
h^s,\,h^{s\prime},\dots \sim \eps_s^2\,,
\ee
and neglect terms that are of order $\eps_s^2$ and higher. Intuitively, this corresponds to and expansion in $\eps_s = \coH/k_S$, where $k_S$ is the wavenumber of small-scale
perturbations. Then, the leading terms in the Einstein
equations are order 1 (such as, for example, the $\partial^2\phi$ term in the 00 component of the Einstein equation in Newtonian gauge).  
Following the same arguments as above, i.e. using that the Einstein
tensor is linear in second derivatives, both
$G^{(\rm nl)F\prime}[h^{s0},h^{s\L}]$ and $G^{(\rm nl)F\prime}[h^{s0},h^{\L}_F]$
appearing in \refeq{EEsLC} are at least order $\eps_s^2$. Thus, for $ k_{S}\gg \coH $ and dropping terms of
order $\eps_s^2$, \refeq{EEsLC} reduces to 
\be
\mbox{CFC, subhorizon:}\quad G_{\mu\nu}^{(\rm lin, SH)}[h^{s \L}]  = 8\pi G
T^{s\L}_{\mu\nu} \,,
\label{eq:EEsLC3}
\ee
where $G_{\mu\nu}^{(\rm lin, SH)}$ is the linear Einstein tensor around the
\emph{local} FLRW background $a_F(\tau_F)$ in the subhorizon limit.  
In other words, \emph{the long-wavelength modes do not appear on the l.h.s. of the CFC
frame Einstein equations in the subhorizon limit.}  
Note that this is not the case in global coordinates, where terms of the form $h^{\L} \partial\partial h^{s0}$ are in general present.  

Let us briefly consider the r.h.s. of the Einstein equations, i.e. the
long-short coupling piece $T^{s\L}_{\mu\nu}$ of the stress-energy tensor perturbation.  The metric enters only without derivatives, so that $h^\L$ does not appear.  Thus, only the fluid perturbations, and $h^{s0},\,h^{s\L}$ can appear
on the r.h.s. of the Einstein equations (moreover, $h^{s\L}$ enters
only through the linear
Taylor term $(\partial T[h^{s0}]/\partial h^{s0}_{\mu\nu}) h^{s\L}_{\mu\nu}$).  

To summarize: in the subhorizon limit of the CFC frame, the long-wavelength 
modes do not
appear in the Einstein equation, but rather enter in the small-scale 
dynamics only through the fluid equations.  If one is interested in
the leading mode-coupling contribution (second order in perturbation
theory), the linear Einstein equations for $h^{s\L}$ are thus sufficient.

\subsubsection{Examples}

Finally, let us briefly discuss two applications.  First, consider 
the coupling of a long-wavelength \textit{tensor} mode $\gamma_{ij}$
to small-scale scalar modes.  We now allow for $k_S$ to be comparable
to $\coH$.  At second order, it is easy to see that
$\gamma_{ij}$ does not enter the $00$, $0i$ and the trace part of the
$ij$ component of the Einstein equations for $h^{s\L}$ [\refeq{EEsLC2}] 
at all for scalar perturbations in conformal-Newtonian gauge.  This 
is because the only 3-tensor available to contract with $\gamma_{ij}$
is $\delta_{ij}$, which gives zero.  The tensor
mode does enter the trace-free part of the $ij$ component of the 
Einstein equations, and thus contributes to an effective anisotropic
stress at second order.  In order to extract the longitudinal
part of these equations, we need to act with
the operator $\partial_i\partial_j/\partial^2$ on the $ij$ Einstein
equation.  As discussed above, this means we need to consider
$\O(x^i)$ and $\O([x^i]^2)$ in the $ij$ Einstein equation.   
We discuss these issues in detail in \refsec{long-tensor} and
\refapp{CFC-inv-spatial-deriv} through \refapp{CFC-compare-Dai}.  

As a second application, consider a long-wavelength \textit{scalar} perturbation.  This is
the focus of the companion paper \cite{CFCpaper2}, and we only give a brief
outlook here.  Let us work in conformal-Newtonian gauge and denote
the long-wavelength potential with $\Phi$, while the short-scale potential
is denoted as $\phi$.  As we will see in \cite{CFCpaper2}, all isotropic effects are absorbed in the locally measurable scale factor $a_F$ and
curvature $K_F$ (correspondingly, the long-wavelength contribution to the
stress energy is absorbed in the effective local ``backgound'' stress energy).  What remains at our disposal is exclusively the purely anisotropic (i.e. trace-free) tidal tensor $(\partial_i \partial_j - (1/3) \delta_{ij} \partial^2) \Phi$ from long-wavelength perturbations.    Note that
this tidal field has the exact same form as the well-known subhorizon limit,
even though here we have not restricted the long-wavelength perturbation
to be subhorizon. 

This means that for an isotropic long-wavelength perturbation, at
second order the linear
Einstein equation for the small-scale modes is \emph{exact on all scales}.  
This can be seen as proof of the ``separate universe'' picture.  
For an anisotropic perturbation, the long-wavelength perturbation can only 
enter the trace-free part of the $ij$ Einstein equation as in the tensor
case, since as before in conformal-Newtonian gauge $\delta_{ij}$ is the only quantity to contract with.   However, this contribution is suppressed
on subhorizon scales $k_S \gg \coH$ in keeping with \refsec{EESH}, so
that for subhorizon small-scale modes, the Einstein equations do not
contain the long-wavelength modes explicitly at all.

\subsection{CFC fluid equations}

In addition to the Einstein equation discussed above, we also need
to consider the fluid equations,
\be
\nabla_\mu T^{\mu\nu} = 0\,,
\ee
which along with the equation of state close the system.  
In the same manner as described above, we can derive the fluid equation
that describes the evolution of the long-short mode coupling contribution
$T^{s\L}$.  Naively, $h^\L_F$ enters into this equation with only one derivative, leading to a vanishing contribution on the central geodesic in CFC.  
However, spatial derivatives of the velocity $\partial_i v^j \sim \partial_i T^j_{\  0}$ appear in the fluid equations.  Thus,
in order to close the hierarchy, we need to take one
further derivative.  Since the velocity itself depends 
on $\partial h^\L$, the equation for $\partial_i v^i$ involves two spatial derivatives
of the metric.  In this way, $h^\L_F$ enters in the small-scale
dynamics in CFC through the fluid equations.  Another way to say this is that
we need to know the velocity at order $x_F$ away from the central geodesic
in order to derive the evolution of the density on the central geodesic.  
This will become apparent in the explicit calculation of \refsec{long-tensor}.

\section{Relating CFC to observations}
\label{sec:relation-FNC}

We now discuss how calculations of observables in the CFC frame
can be related to measurements made by a distant observer.  
This is a crucial step in connecting calculations to actual measurements
(for example, the statistics of galaxy clustering).  Since
the focus of this paper is the CFC frame, and the detailed treatment
of this step depends somewhat on the observable considered, we restrict ourselves to an
outline of the general procedure here, and leave the application 
to specific observables for future work.  

The mapping from CFC to observations made by a distant observer are
often referred to as ``projection effects''.  They correspond to measurable
quantities from the point of view of the distant observer (after all the
events being measured are on the past lightcone of the observer).  However,
they are \emph{not locally measurable} for observers comoving on the
central geodesic of the given CFC frame.  Thus, the CFC frame separates
locally measurable and 
projection contributions to any given observable in an unambiguous way.  
We consider this a key advantage of the CFC formalism.  The projection
effects further separate into two contributions, which we consider in turn.

\subsection{CFC as Lagrangian coordinates and mapping to Eulerian frame}
\label{sec:CFC-lagrange-eulerian}

As we have discussed, the CFC can be constructed for any given geodesic,
which defines the spatial origin of the coordinate system at all times.  
Now consider an ensemble of geodesics of the cosmic fluid.  We can label
each geodesic with a continuous three-dimensional index\footnote{We are considering some finite spatial region here, not necessarily the entire Universe.  Also, recall that we are constructing the CFC for a coarse-grained metric, so that crossing of geodesics can be avoided by choosing an appropriately large smoothing scale.}    $\v{q}$.  This $\v{q}$
is then simply the Lagrangian coordinate of the fluid element moving
along the geodesic labeled by $\v{q}$.  One possible choice for $\v{q}$
is the spatial coordinate in some global coordinate system at a fixed
proper time $t_F =$~const (e.g. at early times).  This shows that the 
CFC frame is related to the standard Lagrangian 
coordinates often used in large-scale structure calculations \cite{Bouchet:1992xz,Buchert:1993df}.  
Since $\v{q}$ usually refers to the position at an early time, where the
initial conditions of cosmological perturbations are set, these coordinates
are especially useful when connecting observables to the statistics of the initial conditions.

\begin{figure}[t!]
\centering
\includegraphics[scale=0.6]{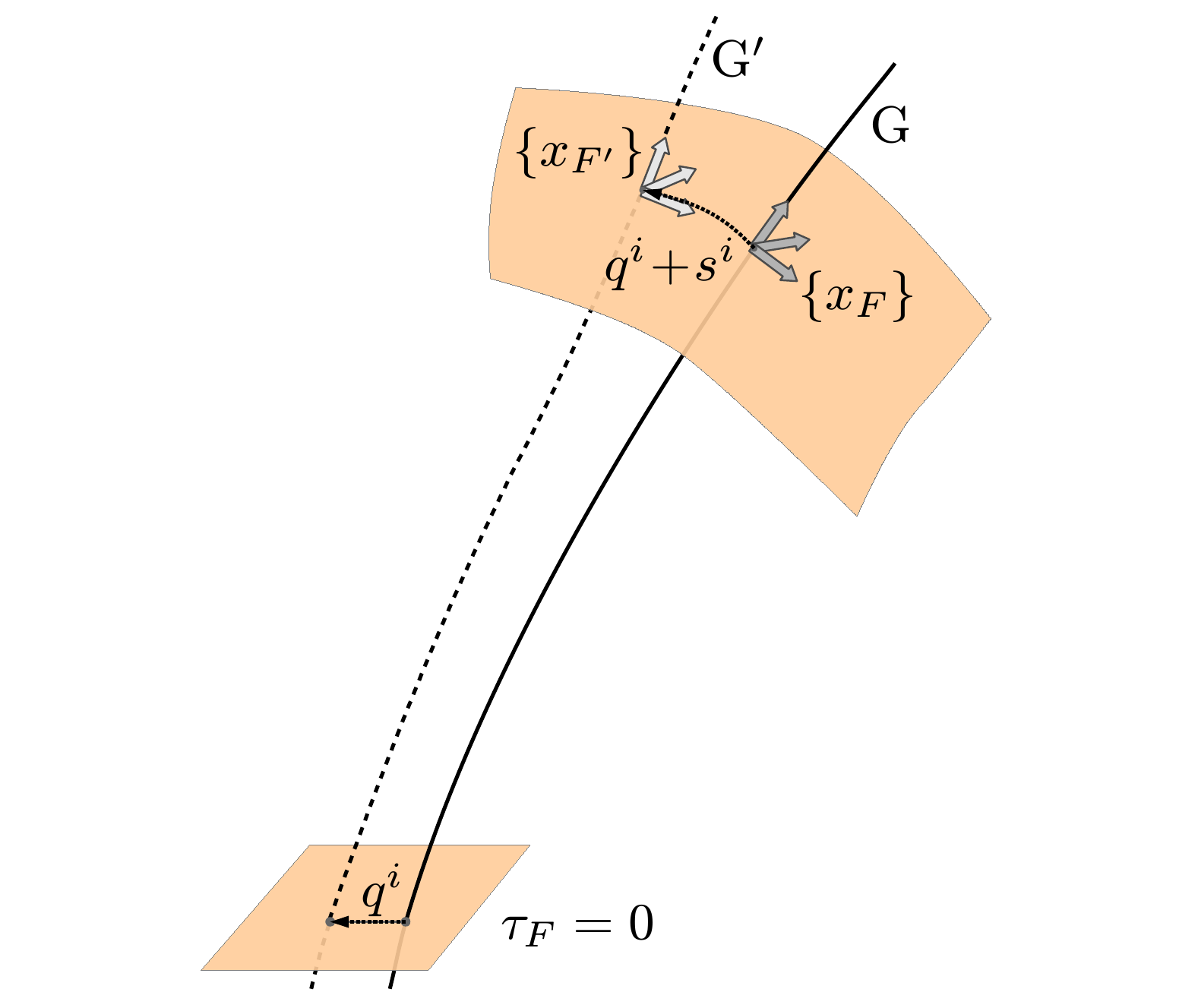}
\caption{Illustration of the mapping between nearby CFC frames $\{x_F\}$, 
$\{ x_{F'} \}$ (tetrads indicated as arrows) constructed for the geodesics $G$, $G'$, respectively
(thick solid and dashed).  The initial separation at $\tau_F = 0$ of the
two geodesics in the $\{ x_F \}$ frame is given by $\v{q}$ (dotted arrow), which at
late times is modified to $\v{q}+\v{s}$ due to the local geodesic deviation [\refeq{shift}].
\label{fig:ELsketch}
}
\end{figure}

However, observations of large-scale structure are made in a Eulerian frame,
which is a frame where coordinates refer to fixed spacetime points rather
than fluid elements.  Thus, we need to transform observables given in the CFC frame to a Eulerian coordinate system.  Suppose we are measuring some observable,
e.g. the density of matter or tracers, in a finite region of space.  For example,
we could be interested in the two-point functions of galaxies or matter
in that region.  Let us assume that this region is defined on a surface
of constant proper time $t_F$.  
We can construct the CFC frame for each geodesic threading
this region, and calculate e.g. the matter density field in that frame (\reffig{ELsketch}).  
However, each geodesic carries a slightly different CFC coordinate system,
while an observer charting the entire region would refer the density measured
at each point to a common ``Eulerian'' coordinate system.  
Of course
these coordinates are arbitrary, but in this context it is most natural
to choose the CFC constructed for the center of the region as our common
coordinate system (see the following subsection).   Let us call this frame $\{ x_F \}$ and the geodesic
for which this frame is constructed G.  We will also refer to it as
``central CFC frame''.    
There are now two options to deal with observables on different nearby geodesics:
\begin{enumerate}
\item Calculate the observable for each CFC frame $\{ x_{F'}\}$ throughout the
region separately, and then transform the observable to the $\{ x_F \}$
frame.
\item Calculate the observable directly in the $\{ x_F \}$ frame, but
away from the geodesic G, i.e. for $x_F^i \neq 0$.  
\end{enumerate}
Either choice can be adopted and has to lead to the same result, since it
is merely a coordinate choice.  In the first option, one has to 
calculate a coordinate transformation between two nearby CFC frames.  
The second option requires keeping at least
one order of $x_F^i$ higher in the Einstein and fluid equations, since we
need to solve them away from the central geodesic of the CFC.  In general,
many new terms beyond those discussed in \refsec{EE} will appear in the CFC
frame away from the geodesic, i.e. at linear order in $x_F^i$.  Some of
these new terms will involve higher derivatives of the long-wavelength
metric perturbation $h^\L$, which should be dropped as they are higher order
in the CFC expansion.  Other terms will correspond to residual gauge
freedom of the CFC construction at higher order in $x_F$.  
Nevertheless, the equivalence of the two approaches
mentioned above shows that, when performing
a consistent expansion in derivatives of $h^\L$, 
\emph{the only physical effect encoded in the Einstein and fluid equations
for $x_F^i \neq 0$ is the effect of the relative displacement 
of neighboring geodesics (geodesic deviation)} at that order in derivatives.  
In practice, the first option is much simpler to pursue (and more physically
transparent), since it corresponds to merely performing 
a coordinate transformation, which turns out to be very simple.  We thus
follow that approach here.

We want to derive the transformation from the CFC constructed for
a geodesic $G'$, which we shall call $\{ x_{F'}\}$, 
to the frame $\{ x_F\}$ around $G$.  We correspondingly denote the local
scale factors as $a_{F'}$ and $a_{F}$, respectively.  
First, let us derive the equation for the
origin of the $\{ x_{F'} \}$ frame in the $\{ x_F\}$ coordinate system, which
we will denote as $s$.    
Since the origins of both frames follow geodesics, this is at leading order
given by the geodesic deviation equation.  Note that the displacement $s$
is not a strictly locally measurable quantity; however it is in principle
observable as long as two observers on $G$ and $G'$ can communicate over the
distance $s$, unlike the ``absolute'' displacement between global Eulerian and
Lagrangian coordinates which is not an observable.   We know that in the absence
of CFC metric corrections, which is equivalent to $\tilde R_F = 0$ on the
central geodesic, the separation vector remains constant since it is a
comoving separation.  The effect we are interested in is thus due to a nonzero
$\tilde R_F$, and we use the geodesic deviation equation for the
conformal metric $\tilde g^F_{\mu\nu}$.  As long as we can assume the scale factor to be the same
$a_F(\tau_F)$ everywhere, this gives the desired result.  Below we will show
that this is justified.  One subtlety arises because
the trajectories we are considering are geodesic with respect to the
full metric, which implies they are \emph{not} geodesic with respect
to the conformal metric (although this would be true if we were considering
null geodesics).  By using the transformation of the Christoffel symbols
under the conformal transformation [\refeq{tilGa}], and rescaling the affine parameter of the geodesic (proper time) $t_F$ to $\tilde\lambda$ defined by $d\tilde\lambda = a_F^{-2} dt_F$, 
the geodesic equation for the full metric can be rewritten as the following equation describing non-geodesic motion with respect to the conformal metric:
\be
\frac{d^2x^\mu}{d\tilde\lambda^2} + \tilde\Gamma^\mu_{\alpha\beta} \frac{dx^\alpha}{d\tilde\lambda}\frac{dx^\beta}{d\tilde\lambda}
= \tilde g_{\alpha\beta} \frac{dx^\alpha}{d\tilde\lambda}\frac{dx^\beta}{d\tilde\lambda} \tilde g^{\mu\gamma} \partial_\gamma \ln a_F\,,
\ee
the deviation from geodesic motion being described by the term on the r.h.s.  
Along the central geodesic, $\tilde\Gamma^F=0$, and this equation simplifies significantly:
\be
\frac{d}{d\tilde\lambda} \hat e^\mu 
= a_F^2 \coH_F \delta^\mu_{\  0}\,,
\ee
where we have defined the tangent vector to the geodesic $\hat e^\mu \equiv dx^\mu/d\tilde\lambda = a_F^2 (e_0)^\mu = a_F (1,0,0,0)$.  
The geodesic deviation equation can easily be generalized to the case where
the curve is not a geodesic (see e.g. the derivation in \cite{SeanCarrollBook}).  Denoting the displacement as $s^\alpha$,
it is given by
\be
\frac{d}{d\tilde\lambda} \left( \frac{d}{d\tilde\lambda} s^\alpha \right) 
= \hat e^\beta \hat e^\gamma (\tilde R_F)^\alpha_{\  \beta\gamma\mu}\, s^\mu
+ s^\gamma \tilde\nabla_\gamma \left( \frac{d}{d\tilde\lambda} \hat e^\alpha \right)\,.
\label{eq:gdev}
\ee
This formidable-looking equation again simplifies greatly in the CFC frame,
since all the covariant derivatives reduce to simple derivatives with respect to $\tau$:
\ba
a_F \left(a_F  s^{\alpha}{}' \right)' =\:& a_F^{2} (\tilde R_F)^\alpha_{\  00\mu}\, s^\mu
+ \delta^\alpha{}_0 s^0 \left(a_F^2 \coH_F\right)'\,,\quad\mbox{or} \vs[8pt]
a_F^{-1}  (a_F s^{\alpha}{}')' 
-\:& \delta^\alpha_{\  0}\, s^0\, \left(\frac{a_F''}{a_F} + \coH_F^2 \right)
=  (\tilde R_F)^\alpha_{\  00\mu}\, s^\mu 
\,.
\label{eq:gdev2}
\ea
Before solving this equation, let us first look at the time component.  
This involves the component $(\tilde R_F)^0_{\  00\mu} = -(\tilde R_F)_{000\mu}$
(recall that the indices are lowered with the conformal metric, $\tilde g_{\mu\nu}^F = \eta_{\mu\nu}$).  By construction of the CFC, this vanishes \emph{at all times:}  since the Christoffel symbols vanish on the central geodesic,
$\tilde R_F$ is only composed of second derivative terms of the metric,
which are non-vanishing only for two spatial derivatives.  We only have one
spatial index at our disposal in $(\tilde R_F)_{000\mu}$ however, so that
this component has to vanish on $G$.  The temporal displacement simply obeys
\be
a_F^{-1}  (a_F s^{0\prime})' - s^0 \left(\frac{a_F''}{a_F} + \coH_F^2 \right) = 0\,.
\ee
This is a second-order homogeneous equation for $s^0$.  It encodes the
fact that $\tau_F = $~const surfaces are not proper time surfaces in CFC.  
Note however that there is no explicit dependence on the long wavelength mode so that
this temporal displacement is not of interest for our purposes.  Similarly,
$s^0$ does not influence the spatial displacement $s^i$ since 
$(\tilde R_F)^i_{\  000}$ vanishes for the same reason.  We can thus
consistently set $s^0$ to zero.

Integrating the spatial component of \refeq{gdev2} yields at first order in $\tilde R_F$
\ba
\label{eq:s-shift-res}
s^i(\tau_F) =\:& \int_0^{\tau_F} \frac{d\tau'}{a_F(\tau')} \int_0^{\tau'} d\tau'' a_F(\tau'') (\tilde R_F)^i_{\  00j}\, q^j 
\vs
=\:& \frac12 \int_0^{\tau_F} \frac{d\tau'}{a_F(\tau')} \int_0^{\tau'} d\tau'' a_F(\tau'') \left(\partial_i \partial_j h^F_{00}\right) q^j\,,
\ea
where $\v{q}$ is the initial displacement.  In other words, $\v{q}$ is the
Lagrangian coordinate assigned to the geodesic $G'$ in the region considered,
where the central geodesic $G$ corresponds to $\v{q} = 0$.  
In the second line we have
expressed the result in terms of the 00 component of the CFC frame metric,
which involves the same components of the Riemann tensor [\refeq{gF-00}].  

Note that an expansion in $\tilde R_F$ is an expansion in the leading corrections
to the CFC metric, in which we retain only linear order terms throughout.   
Hence, a term of order $(\tilde R_F)^2$ is consistently
neglected throughout this paper.  
Now, we note that going away from $G$, the departure of the scale factor
$a_{F'}$ from $a_F$ is, by construction of the CFC, at least of order $(s^i)^2$.  
The tetrad $(e^{F'})$ can be obtained from the tetrad $(e^F)$ for the $\{ x_F \}$
frame by parallel transport along the $x_F^i$ coordinate lines with
respect to the conformal metric $\tilde g_{\mu\nu} = a_F^{-2} g_{\mu\nu}$ (this
is because the CFC requirement is to preserve the form of the conformal metric).  
Again, since $\Gamma^F$ vanish along $G$, the difference between $(e^{F'})$ and
$(e^F)$ is at least order $(s^i)^2$, where as we have seen $s^i$ is of
order $\tilde R_F$.  Thus, at the order in derivatives that
we work in, it is entirely sufficient to only consider the effect of the
centroid shift $s^i$.    

This leads to a trivial transformation of quantities calculated on a
given geodesic within the region to the fixed $\{ x_F \}$ frame (\reffig{ELsketch}).  
For example, let $\rho_{F(\v{q})}$ denote the density (or really any other
local observable) calculated in the local CFC frame 
for the geodesic $G'$ which is labeled by 
the Lagrangian coordinate $\v{q}$ defined above.  Then, at a given time
$\tau_F$, this geodesic is located at a spatial coordinate $\vx_F$ in the central CFC frame
given by
\ba
x_F^i =  q^i + s^i(\v{q},\tau_F) =\:& q^i + \frac12 \int_0^{\tau_F} \frac{d\tau'}{a_F(\tau')} \int_0^{\tau'} d\tau'' a_F(\tau'') \left(\partial_i \partial_j h^F_{00}\right) q^j\,.
\label{eq:shift}
\ea
We can thus relate the density in the local CFC frame $\rho_{F(\v{q})}(\tau_F)$ to the
corresponding result in the central CFC frame $\rho_F(\vx_F,\tau_F)$ through
\be
\label{eq:CFC-lagrangian-eulerian-shift}
\rho_F(x_F,\tau_F) = \rho_{F(\v{q})}(\tau_F)\Big|_{\v{q} = \vx_F - \v{s}(\vx_F,\tau_F)}
= \left[1 - s^i(\vx_F,\tau_F)\, \partial_i\right]\rho_{F(\v{q})}(\tau_F)\Big|_{\v{q}=\vx_F}\,,
\ee
where the derivative acts on the Lagrangian coordinate $\v{q}$.  The first equality
uses the fact that we only need to consider a spatial shift in $\v{q}$, 
while the second equality expands to linear order in the shift.  
This transformation now clearly looks like a transformation from Lagrangian to quasi-local Eulerian coordinates.

\subsection{Photon propagation effects}

We now have an expression for some observable, say $\rho_F(\vx_F,\tau_F)$ 
given in the CFC frame of the central geodesic $G$ of the region considered.  
In order to complete the connection to observations,
we need to relate this CFC frame to the observables of a distant
observer, specifically the photon momentum (frequency or redshift, and 
angular direction) detected at the observer's location.  The mapping from
a local Fermi Normal Coordinate (FNC) frame, constructed for a spatial
region much smaller than the horizon, to photon observables is
well understood; specifically, it was given as ``standard ruler perturbations'' in Ref.~\cite{Schmidt:2012ne} (see also \cite{Jeong:2014ufa} for a review)
at linear order in metric perturbations for a general
perturbed FRW metric.  Thus, as long as the small-scale perturbations 
within the CFC patch are far inside the horizon (which is the most realistic
case), we can make use of these results to relate to observations, by first
transforming CFC to FNC and then mapping FNC to observations.

The CFC are a generalization of the FNC, 
and it is straightforward to derive the mapping from a given CFC frame to FNC.    
Essentially, it only amounts to going from
comoving coordinates $(\tau_F, x_F)$ to physical coordinates $(t_F, \hat x_F)$ 
defined through $dt_F = a_F d\tau_F + \O(x_F^2)$, and $\hat x_F^i = a_F\,x_F^i + \O(x_F^3)$.  
The line element in CFC is of the form
\bea
ds^2 = a^2_F(\tau_F) \left[ \eta_{\mu\nu} + h^F_{\mu\nu} \right] dx^\mu_F dx^\nu_F,
\eea   
where the tidal deviation $h^F_{\mu\nu}$ is $\mathcal{O}[(x^i_F)^2]$ (note
that $h_F$ also includes the contribution from spatial curvature).  

We now first neglect the quadratic terms $h^F_{\mu\nu}$ and construct the usual FNC (following \refsec{FNC}) from the homogeneous flat FLRW metric,
\refeq{dsFLRW} with $K=0$, which yields
\bea
ds^2 & = &  - \left( 1 - \frac{1}{a_F}\frac{d \coH_F}{d t_F}\, \hat r^2_F \right) dt^2_F + \left[ \delta_{ij} + \frac13 \frac{\coH^2_F}{a^2_F} \left( \hat x_{F,i} \hat x_{F,j} - \delta_{ij} \hat r^2_F \right) \right] d\hat x^i_F d\hat x^j_F,
\eea
where $\hat r^2_F= \delta_{ij}\hat x^i_F \hat x^j_F$.  We then restore the tidal deviation $h^F_{\mu\nu}$,
\bea
\label{eq:eq-D2}
ds^2 & = &  - \left( 1 + \frac{1}{a_F}\frac{d \coH_F}{d t_F}\, \hat r^2_F + \frac{1}{a_F^2} h^F_{00} \right) dt^2_F + \frac{2}{a_F^2} h^F_{0i}\, dt_F d\hat x^i_F \nn\\
&& + \left[ \delta_{ij} + \frac13 \frac{\coH^2_F}{a^2_F} \left( \hat x_{F,i} \hat x_{F,j} - \delta_{ij} \hat r^2_F \right) + \frac{1}{a_F^2} h^F_{ij} \right] d\hat x^i_F d\hat x^j_F,
\eea
Factors of $1/a_F^2$ have been added, because $\hat x^i_F$ in FNC refers to the physical separation while in CFC it refers to the ``comoving'' separation.  A few comments on this result are in order.  

First, \refeq{eq-D2} is exactly the FNC metric one would construct directly from a perturbed FLRW spacetime, after inserting the expression for the local expansion rate and acceleration, i.e. $\coH_F/a_F$ and $(1/a^2_F)d\coH_F/d\tau_F$, in terms of background quantities and perturbations, using \refeq{local-Hubble-2} and \refeq{eq-CFC-acceleration}.  
Thus, in addition to extending the regime of validity to beyond the Hubble horizon, the CFC provides an unambiguous separation of the quadratic metric corrections in the standard FNC construction into terms that describe an effective, locally measurable ``background'' expansion (which to zeroth order in $h_{\mu\nu}$ is the actual background), and terms which cannot be absorbed into the background cosmology and therefore describe a departure from a local FLRW universe.   In order to completely prove this statement, we also need to unambiguously define the effective curvature $K_F$ in the CFC frame, which is not made explicit in \refeq{eq-D2}.  We will show this in the companion paper~\cite{CFCpaper2}.  We just mention here that the curvature can be uniquely defined in terms of the spatial CFC metric perturbation $h^F_{ij}$.  This separation of terms into local Friedmann and non-Friedmann terms can help to clarify the physical impact of long-wavelength modes (e.g., \cite{Sherwin:2012nh}).

Second, if our goal is to connect to observations, then there is no need to
insert the expressions for $a_F, \coH_F$ and work out \refeq{eq-D2}.  
As mentioned above, the mapping from the local FNC of the emitting source
to the observer's coordinates of observed redshift and arrival direction of the
detected photons is already known.  Furthermore, for subhorizon 
perturbations (at leading order in $\eps_s$, \refsec{EE}), this mapping
reduces to the well-known lensing and redshift-space distortion effects.  
We then only need to take into account that in CFC, comoving coordinates
are defined with respect to $a_F$ rather than $a$.  Thus, spatial
coordinates need to be rescaled by
\be
x^i_G = \frac{a_F(t_F)}{a(t_F)} x^i_F\,,
\label{eq:cresc1}
\ee
where $t_F$ is the proper time at emission (note that the mapping from observed redshift at observation to proper time at emission is part of what we call the mapping from FNC to observations).  In addition, conformal time intervals
need to be transformed according to
\be
d\tau_G = a^{-1} dt_F = \frac{a_F}{a} d\tau_F\,.
\label{eq:cresc2}
\ee
No such transformation is needed for physical time intervals $dt_G = dt_F$ of course.  
Once the observables are transformed through \refeqs{cresc1}{cresc2}, one can 
then apply the mapping to observations derived in \cite{Schmidt:2012ne}.  

For clarity, let us summarize the three components of the projection effects
necessary to map local CFC-frame observables to measurements by a distant
observer:
\begin{enumerate}
\item Transform from Lagrangian coordinates to a common, ``central'' CFC
frame $\{ x_F \}$ [\refeq{CFC-lagrangian-eulerian-shift}].
\item Rescale comoving coordinates from local scale factor $a_F$ to the
observer's background $a$ [\refeqs{cresc1}{cresc2}].
\item Perform the mapping from local FNC to observations by adding the
ruler perturbations of \cite{Schmidt:2012ne}.
\end{enumerate}

This three-step process may seem fairly complicated.  However, it is
important to emphasize that this is a completely general scheme which works for \emph{any} observable,
i.e. we do not have to re-derive these steps for each new observable
considered.  Second, each of these three steps has a clear physical
interpretation on its own and is very easy to perform in practice.

\section{Application: long-wavelength gravitational waves}
\label{sec:long-tensor}

As specific application of the CFC frame, we consider a FLRW universe perturbed by large-scale gravitational waves, or tensor metric perturbations (see \cite{Masui:2010cz} for previous related work). This amounts to setting $h_{00}=h_{0i}=0$ in \refeq{perturbed-FLRW}, and requiring that $h_{ij}=\gamma_{ij}$ is traceless $\delta^{ij} \gamma_{ij}=0$, and divergence-free $\partial^i \gamma_{ij}=0$ (note that in the notation of the previous section, this is $h^\L$).  The CFC frame can be used to describe the local, measurable impact of the tensor modes over a spatial extent much smaller than the typical wavelength of the tensor modes;  specifically, if the long-wavelength mode is a Fourier mode $k_L$ and the small-scale perturbations considered have a wavenumber $k_S$, then the terms neglected in the CFC frame calculation are suppressed by an additional factor of $k_L/k_S$.  In large-scale structure applications,
this is an excellent approximation for a primordial gravitational wave
background, which is the case we will consider here.

Since tensor modes are traceless, it is easy to see that the local scale
factor $a_F$ [\refeq{aF}] as well as the Hubble rate are the global ones;  
in particular, the global coordinate time $t$ is equal to the proper time
for comoving observers.  
Let us consider small-scale scalar perturbations $\phi,\,\psi$, 
which we describe in conformal-Newtonian gauge (this is $h^s$ in the notation
of \refsec{EE}).  We will assume a matter-dominated universe.  
By construction, the CFC is not restricted to subhorizon scales;  we
thus do not require $k_S \gg \coH$.  For simplicity we will 
restrict to the leading order effect on small scales, and thus insert
the linear solution for the small-scale fluctuations when calculating
the coupling between long wavelength tensor and short scalar modes.  

In the following, we shall drop the subscript ${}_F$ 
from all quantities for simplicity, since we will deal exclusively with CFC frame quantities.  In that frame, the metric becomes 
\bea
\label{eq:tensor-CFC-metric}
ds^2 = a^2(\tau) \left[ \left( - 1 - 2 \phi + h_{00} \right)\, d\tau^2 + 2 h_{0i}\, d\tau dx^i + \left( \left( 1 - 2 \psi \right) \delta_{ij} + h_{ij} \right)\, dx^i dx^j \right]\,,
\eea
where the local influence of the long-wavelength $\gamma_{ij}$ is encoded in the CFC tidal terms $h_{\mu\nu}$,
\bea
\label{eq:tensor-tidal-force-00}
h_{00} & = & \frac12 \left( \coH \gamma'_{kl} + \gamma''_{kl} \right)\, x^k x^l, \\
\label{eq:tensor-tidal-force-0i}
h_{0i} & = & \frac13 \left( \partial_i \gamma'_{kl} - \partial_k \gamma'_{il} \right) \,x^k x^l, \\
\label{eq:tensor-tidal-force-ij}
h_{ij} & = & \frac{1}{6} \coH \left( - \gamma'_{ij}\, \delta_{kl} - \gamma'_{kl}\, \delta_{ij} + \gamma'_{ki}\, \delta_{lj} + \gamma'_{kj}\, \delta_{li} \right)\, x^k x^l \nn\\
&& + \frac16 \left( \partial_k \partial_l\, \gamma_{ij} + \partial_i \partial_j\, \gamma_{kl} - \partial_l \partial_j\, \gamma_{ki} - \partial_l \partial_i\, \gamma_{kj} \right)\, x^k x^l.
\eea
Here, the coefficients multiplying $x^k x^l$ are to be computed along the central geodesic.  Since we assume negligible first-order anisotropic stress, $\gamma_{ij}$ satisfies the equation of motion $\gamma_{ij}''+2\coH \gamma_{ij} - \partial^2 \gamma_{ij}=0$.   

In the remainder of the section we proceed as follows.  We begin with
the conceptually simpler subhorizon case $k_S \gg \coH$ in 
\refsec{CFC-sub-short} (recall that throughout $k_L$ can be arbitrarily small).  
We then describe how the results of the CFC calculation can be related
to observations in \refsec{CFC-distant-obs}.  Finally, we drop the
approximation $k_S \gg \coH$ and solve the full second-order Einstein-fluid
system in \refsec{CFC-allscale-short}.

\subsection{Subhorizon short modes}
\label{sec:CFC-sub-short}

We first restrict the short modes to be in the subhorizon regime $k_S \gg \coH$, but otherwise leave the tensor wavelength fully general.  As discussed in \refsec{EE}, it is sufficient to retain the Einstein equations at linear order in the small-scale modes.  These assume their familiar form in conformal-Newtonian gauge, where we can moreover identify the two potentials $\psi=\phi$ on subhorizon scales.  The Einstein-fluid system for the short-scale fluctuations is most conveniently described by the subhorizon Poisson equation, continuity equation and Euler equation,
\bea
\label{eq:eq-subh-Poisson}
\partial^2 \phi & = & \frac32 \coH^2 \delta, \\
\label{eq:eq-subh-cont}
\delta' + \partial_i \left[(1+\delta) v^i \right] & = & 0, \\
\label{eq:eq-subh-Euler}
v'_i + \coH v_i + \left( v^j\, \partial_j \right) v_i & = & -\partial_i \phi + \frac12 \left( \gamma''_{ij} + \coH \gamma'_{ij} \right) x^j,
\eea
Note that while gravity part is linear, we retain the full nonlinearity in fluid motion. The CFC tidal terms do not enter the first two as we restrict to $x^i=0$. However, the Euler equation must be kept at $\mathcal{O}[x^i]$ to account for a velocity shear $\partial_j v^i$ at $x^i=0$, and the long-wavelength tensor enters at that order by supplying a tidal force $(1/2) \partial_i h_{00}$. We also define the velocity divergence $\theta\equiv \partial_i v^i$.  

In the following, we will solve the system \refeqs{eq-subh-Poisson}{eq-subh-Euler} to second order.  Following our discussion in \refsec{EE}, this is not a necessary restriction:  one could solve the Einstein-fluid system to fully nonlinear order in small-scale modes in the subhorizon regime, e.g. using an N-body simulation,  without having to 
worry about nonlinear terms involving the tensor mode in the Einstein equations.

Following our perturbative approach, the above equations are solved as follows. We  start with solutions for $\delta$, $v$ and $\phi$ linear in scalar and tensor perturbations, at which order scalar and tensor perturbations are decoupled. We then compute corrections to short-wavelength scalar perturbations from their couplings with long-wavelength tensors. With this procedure, the solution, taking $\delta$ for example, is 
\be
\delta = \delta^{(1)} + \delta^{(2)}\,,
\ee
where $\delta^{(1)}$ is the linear solution,  $\delta^{(2)}$ is the 
second-order correction due to mode coupling (denoted by a superscript $s\L$ 
in \refsec{EE}), and correspondingly for the other
variables $\phi,\, v^i$. 

For adiabatic initial conditions, and neglecting the decaying mode for the scalar contribution,\footnote{This is justified since the scalar fluctuations are of much smaller scale than the tensor modes, and thus enter the horizon long before the tensor mode does.} 
the linear solution is given by
\bea
\label{eq:linsol}
\delta^{(1)} = \left( -2 + \frac{2}{3\coH^2} \partial^2 \right) \phi_{\rm ini}, \quad v^{(1)}_i = - \frac{2}{3\coH} \partial_i \phi_{\rm ini} + \frac12 \gamma_{ij}'\, x^j, \quad \theta^{(1)} = - \frac{2}{3\coH} \partial^2 \phi_{\rm ini}\,.
\eea
Throughout matter domination the potential remains constant $\phi^{(1)}(\vx,\tau) = \phi_{\rm ini}(\vx)$. The linear density and velocity divergence only need to be known right on the central geodesic, while the velocity itself needs to be accurate to $\mathcal{O}[x^i]$. It is worth highlighting that in CFC a contribution to $v^i$ linear in the long-wavelength tensor is induced at $\mathcal{O}[x^i]$. At second order, \refeqs{eq-subh-Poisson}{eq-subh-Euler} read
\bea
\label{eq:eq-7.10}
\partial^2 \phi^{(2)} - \frac32 \coH^2 \delta^{(2)} & = & 0, \\
\label{eq:eq-7.11}
\delta^{(2)\prime} + \theta^{(2)} & = & - \partial_i \left[ \delta^{(1)} v^{(1)i} \right], \\
\label{eq:eq-7.12}
\theta^{(2)}{}' + \coH\, \theta^{(2)} + \partial^2 \phi^{(2)} & = & - \left(\partial_i v^{(1)j} \right) \left(\partial_j v^{(1)i} \right) - \left( v^{(1)i} \partial_i\, \partial_j\, v^{(1)j} \right).
\eea
When linear solutions are inserted on the right hand sides, we only have to keep mixed quadratic terms proportional to linear short-scale scalar perturbation multiplying linear long-wavelength tensor, as we focus on tensor-scalar coupling. An equation for $\delta^{(2)}$ can be then derived by eliminating $\theta^{(2)}$ and $\partial^2 \phi^{(2)}$. When combining equations, we take time derivatives, but additional spatial derivatives are avoided in order not to mix up different orders in $x^i$ in CFC. We find
\bea
\delta^{(2)}{}'' + \coH \delta^{(2)}{}' - \frac32 \coH^2 \delta^{(2)} & = & - \frac{2}{3\coH} \gamma_{ij}'\, \partial^i \partial^j \phi_{\rm ini}\,.
\eea
Before integrating this equation, we need to specify initial conditons.  
This corresponds to the physical coupling of long-wavelength tensor modes
with short-scale scalar modes when the latter cross the horizon during
inflation.  Since the tensor modes have long exited the horizon at that
time, we do not expect the scalar modes to show any physical coupling
with these long-wavelength tensor modes.  Indeed, \cite{Pajer:2013ana}
showed that the squeezed-limit tensor-scalar-scalar bispectrum in 
single-field inflation vanishes when evaluated in the CFC frame.  
Thus, the initial condition is $\delta^{(2)}(\tau_{\rm ini}) = 0 = \delta^{(2)\prime}(\tau_{\rm ini})$.  The solution reads
\bea
\label{eq:d2CFC}
\delta^{(2)}(\tau) & = & - \frac{2}{3\coH^2} \mathcal{S}_N (k_L,\tau) \left[ \gamma^{ij}_{\rm ini}\,\partial_i \partial_j \phi_{\rm ini} \right] = - \mathcal{S}_N (k_L,\tau) \gamma^{ij}_{\rm ini} \left[\frac{\partial_i \partial_j}{\partial^2} \delta^{(1)}_{\rm sc}(\tau) \right] \vs
\mathcal{S}_N (k_L, \tau) & = & \frac25 \int_0^{k_L \tau} dx
\frac{d\mathcal{T}_\gamma(x)}{dx} \left[1 - \left(\frac{x}{k_L\tau}\right)^5\right]\,.
\eea
Here $\gamma^{ij}_{\rm ini}$ is the initial, superhorizon amplitude of the tensor mode and $\mathcal{T}_\gamma(k \tau) = 3 j_1(k \tau)/(k\tau)$ is the tensor linear transfer function.   In the second equality, we have used that the linear density constrast in the synchronous-comoving gauge $\delta^{(1)}_{\rm sc}$ is related to the Newtonian gauge potential in Einstein-de Sitter universe through $\delta^{(1)}_{\rm sc} = (2/3\coH^2) \partial^2 \phi_{\rm ini}$. 

\subsection{Relating to a distant observer}
\label{sec:CFC-distant-obs}

\refeq{d2CFC} describes modified clustering due to long-wavelength gravitational waves as a local observer would see at her location. A distant observer, however, would measure how density varies over a finite region of space. If the distant observer wants to compute the power spectrum of the apparent clustering, she would need the leading order correction to account for deviation from the CFC central geodesic ($x^i \neq 0$) in which a result like \refeq{d2CFC} is derived. Fortunately, as already discussed in \refsec{CFC-lagrange-eulerian}, this does not require an independent calculation that goes to the next order in $x^i$, but merely a spatial shift as in \refeq{CFC-lagrangian-eulerian-shift}. 

Applying \refeq{tensor-tidal-force-00} to \refeq{s-shift-res}, we find the shift vector to be
\bea
s^i = \frac12 q^j \int^\tau_0 \frac{d\tau'}{a(\tau')} \int^{\tau'}_0 d\tau''\, a(\tau'') \left[ \gamma_{ij}'' + \coH \gamma_{ij}' \right] = - \frac12 \gamma^{\rm ini}_{ij} q^j \left( 1 - \mathcal{T}_{\gamma} \right),
\eea
To leading order, we replace the Lagrangian coordinate $q^i$ with $x^i$ and apply \refeq{CFC-lagrangian-eulerian-shift} to find the quasi-local Eulerian density up to second order
\bea
\label{eq:eq-delta-quasi-euler}
\delta_{\rm E} & = & \delta^{(1)} + \frac12 \left( 1 - \mathcal{T}_{\gamma} \right) \gamma^{ij}_{\rm ini}\, x_i\, \partial_j \delta^{(1)} - \mathcal{S}_N (k_L)\, \gamma^{ij}_{\rm ini} \left[ \frac{\partial_i \partial_j}{\partial^2} \delta^{(1)}_{\rm sc} \right]. 
\eea
On subhorizon scales, which we have assumed throughout this section,
$\delta^{(1)}_{\rm sc} = \delta^{(1)}$ and we have complete agreement with 
the expression obtained in \cite{Schmidt:2013gwa} [Eq.~(55) there]. 

Turning to the comparison with \cite{Dai:2013kra}, the second term in \refeq{eq-delta-quasi-euler} manifests itself as the first term in Eq.~(34) there;  note that in that calculation the bispectrum in comoving gauge \cite{Maldacena:2002vr} was used as initial condition in the second order calculation, leading to the factor of $(1-\mathcal{T}_\gamma)$ of this term.
The third term is equal to the second term in Eq.~(34) of \cite{Dai:2013kra}~\footnote{The sign of this term in Eq.~(34) of \cite{Dai:2013kra} is a typo.}.  The third term in the latter is due to the mapping from local FNC to observations, i.e. the third contribution in the list of projection effects at the end of \refsec{relation-FNC} (the second in that list being absent in this case, since $a_F = a$).  If we include this projection effect as well, in the notation of \cite{Dai:2013kra}, we obtain for the anisotropic distortion of the density power spectrum [defined in Eqs.~(31)--(32) there]
\be
\mathcal{Q}_{ij} = - \frac12 \frac{d \ln P_{\delta}}{d\ln k} \left( 1 - \mathcal{T}_{\gamma} \right) \gamma_{ij}^{\rm ini} - 2 \mathcal{S}_N(k_L,\tau) \gamma_{ij}^{\rm ini}
- \frac{d\ln P_\delta}{d\ln k}\left(\frac12 \gamma_{ij} + \partial_{(i} \Delta x_{j)} - \frac13 \delta_{ij}\, \partial_k \Delta x^k\right)\,,
\ee
where $P_\delta$ is the linear matter power spectrum.  
Note that when performing a Fourier transform of $\delta_{\rm E}$, the explicit dependence on $x^i$ present in CFC turns into a derivative with respect to $k_S$ (see e.g. App. F of Ref.~\cite{Dai:2013kra}).  

The CFC thus automatically provides a clear physical interpretation of each of these
contributions, which makes it considerably easier to check the calculation for correctness.  
Indeed, even in the very restricted case of long-wavelength tensors coupling to short-wavelength scalars, gauge artifacts can be very difficult to control when calculating in global coordinates at second order. 

\subsection{Short modes on all scales}
\label{sec:CFC-allscale-short}

Since CFC is not restricted to subhorizon distances, we now generalize beyond the limit $k_S \gg \coH$ assumed in \refsec{CFC-sub-short}.  We derive the fully general relativistic result for the second-order density measured on the central geodesic.  Compared to the subhorizon case, this does involve some additional subtleties in the CFC resulting from the longitudinal part of the trace-free
$ij$ component of the Einstein equations.  First, while the $00$, $0i$, and trace
component of the Einstein equation can be taken to be linear as in the
subhorizon case, the trace-free component contains terms of the type
$\partial^2 \gamma_{ij}$ and $\coH \gamma'_{ij}$ multiplied by $\phi,\,\psi$.  
Second, unlike the other Einstein equations, we need to act with a spatial
derivative operator in order to extract the longitudinal part of the
trace-free component.

It is convenient to use the $00$-Einstein equation, the divergence of the $0i$-Einstein equation (with {\it linearised} Einstein tensor), supplemented with the (nonlinear) equation of energy conservation, as well as the divergence of the momentum conservation, 
\bea
\partial^2 \psi - 3 \coH \psi' - 3 \coH^2 \phi - \frac32 \coH^2 \delta & = & 0, \\
- \partial^2 \left( \psi' + \coH \phi \right) - \frac32 \coH^2 \partial_i \left[ \left(1+\delta\right) v^i \right] & = & 0, \\
\delta{}' + \partial_i \left[ \left(1+\delta\right) v^i \right] - 3 \left(1+\delta\right) \psi{}' & = & 0, \\
\theta' + \coH\, \theta + \partial^i \left[ \left( v^j \partial_j \right) v_i \right] + \partial^2 \phi & = & 0.
\eea
According to \refsec{EE}, these equations are valid at $x^i=0$.  
At first sight, these are four equations for four unknowns $\phi,\,\psi,\,\delta,\,\theta$, and should thus suffice to solve the system.  However, 
this is not correct in CFC due to the breaking of translation invariance:
in order to solve the system, we would have to take inverse spatial derivatives, which cannot be consistently done at a single point but requires higher orders
in $x^i$ to be included as well (see \refapp{CFC-inv-spatial-deriv}).  
One can however close the system without
taking additional (inverse) spatial derivatives of any of the above
equations by including the longitudinal
part of the trace-free Einstein equations, as we will see below.

Note that we now have to distinguish between the two scalar potentials $\phi$ and $\psi$, since at second order they are no longer equal (although their difference is negligible on subhorizon scales $k_S \gg \coH$).  
We therefore introduce $\Gamma^{(2)}\equiv \psi^{(2)}-\phi^{(2)}$.  
Physically, this stems from an effective anisotropic stress induced at second order by linear perturbations.  
Following the perturbative procedure adopted in \refsec{CFC-sub-short} [where the linear solutions still given by \refeq{linsol}], we obtain equations for second-order perturbations,
\bea
\label{eq:eq-7.22}
\partial^2 \psi^{(2)} - 3 \coH \psi^{(2)}{}' - 3 \coH^2 \phi^{(2)} - \frac32 \coH^2 \delta^{(2)} & = & 0, \\
\label{eq:eq-7.23}
- \partial^2 \left( \psi^{(2)}{}' + \coH \phi^{(2)} \right) - \frac32 \coH^2 \theta^{(2)} & = & \frac32 \coH^2 \partial^i \left[ \delta^{(1)} v^{(1)}_i \right], \\
\label{eq:eq-7.24}
\delta^{(2)}{}' + \theta^{(2)} - 3 \psi^{(2)}{}' & = & - \partial^i \left[ \delta^{(1)} v^{(1)}_i \right] + 3 \delta^{(1)}\,\psi^{(1)}{}', \\
\label{eq:eq-7.25}
\theta^{(2)}{}' + \coH \theta^{(2)} + \partial^2 \phi^{(2)} & = & - \left( \partial^i v^{(1)}_j \right) \left( \partial^j v^{(1)}_i \right) -  v^{(1)}_i \partial^i \partial^j v^{(1)}_j.
\eea
Again, on the right, only quadratic source terms proportional to the product of linear scalar modes and linear tensor modes need to be kept. {\it Without taking additional spatial derivatives}, we can combine \refeq{eq-7.23} and \refeq{eq-7.25} to find an equation for $\partial^2 \psi^{(2)}$ along $G$,
\bea
\label{eq:eq-7.26}
\left( \partial^2 \psi^{(2)} \right)'' + 3 \coH\,\left( \partial^2 \psi^{(2)} \right)' & = & - \coH \gamma_{ij}' \left( \partial^i \partial^j \phi^{(1)} \right) + \coH\,\partial^2 \Gamma^{(2)}{}'.
\eea 
Alternatively, we can combine all equations \refeqs{eq-7.22}{eq-7.25} to obtain an equation for $\psi^{(2)}$ itself along $G$, again without taking spatial derivatives,
\bea
\label{eq:eq-7.27}
\psi^{(2)}{}'' + 3 \coH\,\psi^{(2)}{}' & = & \frac13 \partial^2 \Gamma^{(2)}  + \coH\, \Gamma^{(2)}{}'.
\eea
Comparing \refeq{eq-7.26} and \refeq{eq-7.27} illustrates the point made
above: we cannot simply take the Laplacian of the latter to obtain the
former, since we have neglected terms of $\mathcal{O}[x^i]$ and $\mathcal{O}[(x^i)^2]$ in
both equations.   

Let us now turn to the equation for $\partial^2\Gamma^{(2)}$ at $x^i=0$, 
obtained from the longitudinal part of the trace-free $ij$ component of
the Einstein equations, that is,
\be
\left(\frac{\partial^i\partial^j}{\partial^2} - \frac13 \delta^{ij}\right)
\left[G_{ij} - 8\pi G\,T_{ij} \right] = 0\,.
\ee  
In order to consistently evaluate this equation including the derivative operator acting on $G_{ij}$, we need to explicitly evaluate $G_{ij}$ to second-order (retaining terms of scalar perturbations multiplying tensor perturbations) and up to $\mathcal{O}[(x^i)^2]$ (since any power of $x^i$ from $h^F_{\mu\nu}$ can be converted into an inverse derivative acting on the short-scale perturbation, see \refapp{CFC-inv-spatial-deriv}).  We then set $x^i=0$ at the very end.  This is technically somewhat involved, but can be done as detailed in \refapp{CFC-2nd-aniso-stress}.  We finally obtain on $G$
\bea
\label{eq:Gamma-2nd-res}
\partial^2 \Gamma^{(2)} & = & \frac32 \left( \gamma_{ij}'' + 2\coH \gamma_{ij}' + \partial^2 \gamma_{ij} \right) \left[ \frac{\partial^i \partial^j}{\partial^2} \phi^{(1)} \right]\,.
\eea 
Apart from $\partial^2 \Gamma^{(2)}$, we also see that $\Gamma^{(2)\prime}$
enters \refeq{eq-7.27}.  However, unlike $\partial^2 \Gamma^{(2)}$, 
terms of $\Gamma^{(2)}$ and $\Gamma^{(2)\prime}$ are negligible at the order in CFC
expansion that we work in.  To see this, note that the solution for
$\Gamma^{(2)}$ will involve more inverse derivatives acting on $\phi^{(1)}$. More specifically, a net negative power of derivatives act on $\phi^{(1)}$. These terms cannot be determined unambiguously from the CFC metric up to $\O[(x^i)^2]$, as they are sensitive to $\O[(x^i)^3]$ terms in the CFC metric, which we have
neglected throughout. They are also affected by the residual gauge freedom \refeq{residual-gauge-hFij} up to $\O[(x^i)^2]$. Physically, this means that these terms are beyond the leading local effect of long modes in terms of an expansion in $k_L/k_S$.Thus, we neglect the term $\coH\Gamma^{(2)\prime}$ in \refeq{eq-7.27}.  
We then integrate \refeqs{eq-7.26}{eq-7.27} to find
\bea
\partial^2\psi^{(2)} & = & - \mathcal{S}_N(k_L) \gamma^{ij}_{\rm ini} \left( \partial_i \partial_j \phi_{\rm ini} \right) + 3 \mathcal{S}_N(k_L)\left( \partial^2 \gamma^{ij}_{\rm ini} \right) \left( \frac{\partial_i \partial_j}{\partial^2} \phi_{\rm ini} \right), \\
\psi^{(2)} & = & - \mathcal{S}(k_L) \gamma^{ij}_{\rm ini} \left( \frac{\partial_i \partial_j}{\partial^2} \phi_{\rm ini} \right),
\eea
where
\bea
\label{eq:eq-SN-S}
\mathcal{S}(k_L) & = & \mathcal{S}_N (k_L) + 1 - \mathcal{T}_\gamma (k_L),
\eea
as introduced in~\cite{Dai:2013kra}. 
From \refeq{eq-7.22}, we then find the density on $G$ at second order $\delta_F=\delta^{(1)}+\delta^{(2)}$ to be
\bea
\label{eq:eq-7.31}
\delta_F & = & \delta^{(1)} + \frac{2}{3\coH^2} \partial^2 \psi^{(2)} - 2 \left( \frac{1}{\coH} \psi^{(2)}{}' + \psi^{(2)} \right)  + 2 \Gamma^{(2)}  \nn\\
& = & \left( -2 + \frac{2}{3\coH^2} \partial^2 \right) \phi_{\rm ini} - \frac{2}{3\coH^2} \mathcal{S}_N (k_L) \gamma^{ij}_{\rm ini} \left( \partial_i \partial_j \phi_{\rm ini} \right) +  \frac{2}{\coH^2} \mathcal{S}_N (k_L)\left( \partial^2 \gamma^{ij}_{\rm ini} \right) \left( \frac{\partial_i \partial_j}{\partial^2} \phi_{\rm ini} \right) \nn\\
& & + 2\, \gamma^{ij}_{\rm ini} \left( \frac{\partial_i \partial_j}{\partial^2} \phi_{\rm ini} \right) \left[ \frac{1}{\coH} \frac{d \mathcal{S}(k_L)}{d\tau} + \mathcal{S}(k_L) \right]\,. 
\eea
We have neglected the term involving $\Gamma^{(2)}$ in the first line because the leading result for it involves a net negative powers of derivatives acting on $\phi^{(1)}$. 
\refeq{eq-7.31} shows that the long-wavelength tensor modes couple
to small-scale scalar modes only through $\partial_i\partial_j\phi$ in
conformal-Newtonian gauge, which is directly related to the density
perturbation in synchronous gauge.

As shown in \refapp{CFC-compare-Dai}, the full general relativistic result for the density \refeq{eq-7.31} agrees with~\cite{Dai:2013kra}, where the computation is carrried out in global coordinates, once a coordinate shift relating the two gauges is taken into account.\footnote{Note also that \cite{Dai:2013kra} neglected the term corresponding to $\coH \partial^2 \Gamma^{(2)\prime}$ since it is numerically suppressed by $(k_L/k_S)^2$ after the tensor equation of motion is applied. However, this term is an unambiguous prediction from the CFC metric up to $\O[(x^i)^2]$ and should be kept. The results agree when this term is included.}  Note that when $k_S$ is of order $\coH$, the second order density 
given in \refeq{eq-7.31} which is calculated in conformal-Newtonian gauge 
is not an observable, hence results in different
gauges do not need to agree.  However, as we show in \refapp{CFC-compare-Dai},
the conformal-Newtonian gauge in CFC, and the global conformal-Newtonian gauge
are at lowest order related by a simple coordinate shift related to the
initial tensor mode amplitude.  Thus, the agreement in the second order
density is a nontrivial check of the two different calculations.  
Finally, it is worth pointing out that the case $k_S \sim \coH$ is
not really of practical interest, since it implies that the tensor mode
with $k_L \ll k_S$ is far outside the horizon, suppressing its
observable effects.

\section{Summary and conclusions}
\label{sec:concl}

In this paper we have presented a systematic way to construct Conformal Fermi Coordinates (CFC). These are the coordinates of a free-falling observer that describes her neighborhood as an FLRW space up to corrections that grow with the square of the distance from her worldline, as summarized in \refeq{gF-00}, \eqref{eq:gF-0i} and \eqref{eq:gF-ij}. The coordinates are defined for any, arbitrarily inhomogeneous and anisotropic spacetime, as long as the geodesic maintains a finite distance to any singularities.  In cosmological applications, one typically wants to define CFC with respect to some long-wavelength perturbation. The coordinates are then valid only in a region that is much smaller than the wavelength of that perturbation; specifically, we have shown that higher order corrections are suppressed by powers of the conformal Riemann tensor and its spatial derivatives, $(\partial^m \tilde R_F)^n x_F^{n(m+2)}$ with $m+n>1$.  The key advantage that makes CFC very useful for cosmological computation is that they are \textit{not} restricted to subhorizon scales, contrary to what is the case for standard Fermi Normal Coordinates (FNC).   Indeed, the standard coordinates chosen by cosmologists to describe their observations are the CFC constructed for our entire observable universe.  There are three major perks of using CFC. First, CFC precisely describe what a local observer would measure, eliminating the issue of gauge artifacts at nonlinear order from the start.  Second, the computation of non-linear mode coupling is simplified considerably with respect to standard gauge choices employed in cosmological perturbation theory (e.g., Newtonian or comoving gauges) when $k_S \gg \coH$ (the case of greatest practical relevance).   
We have shown 
(\refsec{EE}) that in that case it is sufficient to retain the linear Einstein
equations for the small-scale modes.  
Finally,
the CFC allow for an unambiguous connection to observations via a well-defined
coordinate transformation (from local to global scale factor and from Lagrangian to Eulerian coordinates), and gauge-invariant projection effects which have been calculated previously.  Moreover, they cleanly separate
different physical contributions of the long-wavelength mode due to genuine 
local gravitational effects, local geodesic deviation, and photon propagation
(projection) effects.  

For example, Ref.~\cite{Pajer:2013ana}, where CFC were first introduced, showed in a very simple calculation that there is no modulation of small-scale perturbations by large-scale potential fluctuations in single-field inflation in the attractor regime.  That is, local primordial non-Gaussianity is completely absent.  Similarly, Ref.~\cite{Pajer:2013ana} showed that there is no physical coupling of large-scale tensor modes to small-scale perturbations in single-field inflation.  This corresponds to the absence of $\O(1)$ and $\O(k_L)$ effects of long-wavelength metric perturbations of wavenumber $k_L$ in the regime $k_L/\coH \ll 1$.  

The advantages of the CFC just described now also allow us to calculate the leading \emph{observable} effect of long-wavelength modes in single-field inflation, which starts at $(k_L/k_S)^2$.  As an explicit example of this, we have computed the effect of a long-wavelength tensor mode on a short-wavelength scalar perturbation (the result is given in \refeq{eq-7.31}).  For a subhorizon scalar perturbation, this had already been computed in \cite{Schmidt:2013gwa} using CFC,
and we recover the same results in that limit here.  
A calculation in global coordinates was presented in \cite{Dai:2013kra},
which in the calculation of the second order density did not assume 
subhorizon small-scale modes (the connection to observations was however still
done assuming $k_S \gg \coH$).  Although the calculation for horizon-scale
small-scale modes in CFC involves some subtleties, it can be done and we
show that we find perfect agreement with the global calculation once
the proper gauge transformation (a simple coordinate shift) is taken into
account.  
The conclusion is that the quantity capturing the mode coupling effect on all scales is the second spatial derivative of the potential $\phi$ in Newtonian gauge, which is directly related to the density perturbation in synchronous-comoving gauge.  

Getting rid of all this apparent nonlinearity of gravity might seem like a 
bit of a miracle.  However there are several reasons why this works.  First,
by isolating local observables, we remove a large number of terms that are merely gauge modes.  
Second, we are relying on a separation in scale between long- and short-wavelength 
modes, so that the long-wavelength modes are almost constant over the scales
relevant for the short-wavelength modes (specifically, the leading tidal 
contribution of the long-wavelength mode is approximated as spatially constant).  Finally, we are making use of the fact
that while the Einstein equations are linearized in \emph{perturbations}, the
background is solved to fully nonlinear order.  This ensures that all 
effects of a long-wavelength mode that can be captured by an effective background $a_F(\tau_F)$
are solved for exactly, i.e. at fully nonlinear order.  This is related to
what is commonly called the ``separate universe'' conjecture, which will
be studied in detail in the companion paper \cite{CFCpaper2}.

Finally, it is important to remember the limitations of our approach. 
By construction, the CFC only capture gravitational effects of the long-wavelength mode.  Thus, they are valid as long as non-gravitational interactions are irrelevant on the scale of the long-wavelength mode, i.e. as long as the long mode is outside of the sound horizon of any non-gravitational interactions.   
Hence certain computations, specifically those where the non-gravitational physics is nonlocal on the scale of the horizon $\coH^{-1}$, must be done in global coordinates. A specific example would be the effect of long-wavelength tensor modes on scalar perturbations before recombination.

\acknowledgments

We would like to thank Tristan Smith, Simon White, and Matias Zaldarraga for discussions.  
Tensorial algebra are partially performed with {\tt xPand}~\cite{Pitrou:2013hga}. E.P. is supported by the D-ITP consortium, a program of the Netherlands organization for scientific research (NWO) that is funded by the Dutch Ministry of Education, Culture and Science (OCW). L.D. was supported by the John Templeton Foundation as a graduate research associate.


\appendix

\section{Evaluating $a_F$ and its derivatives}
\label{app:compute-conformal-connection}

In this section we give coordinate-free expressions for the first and second 
derivatives of $a_F$, which are necessary in order to derive the explicit
transformation from some coordinate frame to CFC.  These derivatives
are already fixed by the fact that $a_F = a_F(\tau_F)$ in CFC and by 
the form of the metric \refeq{CFC-requirement}.  

We start with the local comoving expansion rate along the central geodesic,
\bea
\coH_F(P) = \frac{d\ln a_F(P)}{d\tau_F} = \frac{dx^\mu_P}{d\tau_F} \left(\nabla_\mu \ln a_F \right)_P = a_F(P) (e_0)^\mu_P \left( \nabla_\mu \ln a_F \right)_P\,.
\eea 
Since the gradient of $a_F$ along the central geodesic points in the tangent direction of the central geodesic, we can solve this for the gradient,
\bea
\label{eq:aF-gradient}
\left( \nabla_\mu \ln a_F \right)_P = - \frac{\coH_F(P)}{a_F(P)} (e_0)_{\mu,P}\,.
\eea
This determines the tensor of shift \refeq{Ctensor}.  In order to determine the coordinate transformation up to third order in $x$, we also need derivatives of the conformal Christoffel symbols,  $\pdtilGa{\mu}{\alpha}{\beta}{\gamma}=\pdGa{\mu}{\alpha}{\beta}{\gamma}-\pdCten{\mu}{\alpha}{\beta}{\gamma}$, again
evaluated on the central geodesic. This involves the derivative
\bea
&& \pdCten{\mu}{\alpha}{\beta}{\gamma} = \delta^\mu_\alpha \,\nabla_\gamma \nabla_\beta \ln a_F + \delta^\mu_\beta\, \nabla_\gamma \nabla_\alpha \ln a_F - g_{\alpha\beta}\, g^{\mu\lambda} \nabla_\gamma \nabla_\lambda \ln a_F \nn\\
&& - \left( g_{\alpha\beta,\gamma}\, g^{\mu\lambda} + g_{\alpha\beta}\, g^{\mu\lambda}{}_{,\gamma} \right) \nabla_\lambda \ln a_F + \left( \delta^\mu_\alpha \Ga{\nu}{\gamma}{\beta} + \delta^\mu_\beta \Ga{\nu}{\gamma}{\alpha} - g_{\alpha\beta} g^{\mu\lambda} \Ga{\nu}{\gamma}{\lambda} \right)\, \nabla_\nu \ln a_F\,.
\eea  
Note that the metric \refeq{CFC-requirement} implies that the second
derivatives $\nabla_\alpha \nabla_\beta \ln a_F$ appearing in the first line
are given by the corresponding result for a homogeneous FLRW metric
(spatial curvature does not enter at this order).  This is because the
conformal Christoffel symbols quantifying the departure of the metric
from FLRW vanish along the central geodesic.  Writing this in
a coordinate-free form finally yields
\bea
\label{eq:dadblnaF}
(\nabla_\alpha \nabla_\beta \ln a_F)_P = \left[ \frac{1}{a^2_F(P)} \frac{d\coH_F(P)}{d\tau_F} - 2 \left( \frac{\coH_F(P)}{a_F(P)} \right)^2 \right] (e_0)_{\alpha,P} (e_0)_{\beta,P} - \left( \frac{\coH_F(P)}{a_F(P)} \right)^2 g_{\alpha\beta}.\qquad
\eea

\section{Coordinate transformation}
\label{app:coord}

In this section we derive the transformation from some global coordinate
system to CFC explicitly up to third order in $x_F^i$, continuing from
\refeq{alpha1}.  Higher-order coefficients $\alpha^\mu_n$ are recursively computed using \refeq{conformal-geodesic-eq}.  
At second and third order, we obtain respectively
\bea
\alpha^\mu_2 = \frac12 \left. \frac{d^2x^\mu}{d\lambda^2} \right|_P = - \frac12  \left. \tilGa{\mu}{\alpha}{\beta} \frac{dx^\alpha}{d\lambda} \frac{dx^\beta}{d\lambda} \right|_P = -\frac12 a^2_F(P) \left( \tilGa{\mu}{\alpha}{\beta} \right)_P (e_i)^\alpha_P (e_j)^\beta_P\, x^i_F x^j_F,
\eea  
and
\bea
\alpha^\mu_3 & = & \frac16 \left. \frac{d^3 x^\mu}{d\lambda^3} \right|_P = - \frac16 \frac{d}{d\lambda} \left( \tilGa{\mu}{\alpha}{\beta} \frac{dx^\alpha}{d\lambda} \frac{dx^\beta}{d\lambda} \right)_P \nn\\
& = & - \frac16 a^3_F(P) \left( \partial_\gamma \tilGa{\mu}{\alpha}{\beta} - 2 \tilGa{\mu}{\sigma}{\alpha} \tilGa{\sigma}{\beta}{\gamma} \right)_P \left(e_i\right)^\alpha_P  \left(e_j\right)^\beta_P \left(e_k\right)^\gamma_P\, x^i_F x^j_F x^k_F.
\eea
The transformation from the global coordinates into the CFC, expanded to third order in $x^i_F$, is then given by
\bea
\label{eq:CFC-coordinate-shift}
\delta x^\mu & \equiv & x^\mu(Q) - x^\mu(P) = a_F(P) (e_i)^\mu_P\,x^i_F - \frac12 \left( \tilGa{\mu}{\alpha}{\beta} \right)_P \left[ a_F(P) \right]^2 \left(e_i\right)^\alpha_P \left(e_j\right)^\beta_P \, x^i_F x^j_F \nn\\
&& - \frac16 \left( \partial_\gamma \tilGa{\mu}{\alpha}{\beta} - 2 \tilGa{\mu}{\sigma}{\alpha} \tilGa{\sigma}{\beta}{\gamma} \right)_P \left[ a_F(P) \right]^3 \left(e_i\right)^\alpha_P  \left(e_j\right)^\beta_P \left(e_k\right)^\gamma_P\, x^i_F x^j_F x^k_F + \mathcal{O}\left[ \left(x^i_F\right)^4 \right]\,.
\eea

\subsection*{Transforming the metric}

We now show how the metric transformation can be calculated explicitly 
using the coordinate shift derived above.  
Coordinate shifts up to $\mathcal{O}[(x^i_F)^3]$ as in \refeq{CFC-coordinate-shift} fix the Jacobian matrix $\partial x^\mu / \partial x^\nu_F$ to $\mathcal{O}[(x^i_F)^2]$, corresponding to the leading corrections in CFC frame from the background FLRW metric.  Here and throughout, $P$ denotes a point on the central geodesic.  
Evaluating $\partial x^\mu / \partial \tau_F$ requires the derivatives along the central geodesic,
\bea
\frac{1}{a_F(P)} \frac{d x^\mu(P)}{d\tau_F} & = & (e_0)^\mu_P, \\
\frac{1}{a_F(P)} \frac{d a_F(P)}{d\tau_F} & = & \coH_F(P), \\
\frac{1}{a_F(P)} \frac{d (e_i)^\mu_P}{d\tau_F} & = & - \left(\Ga{\mu}{\alpha}{\beta}\right)_P (e_0)^\alpha_P (e_i)^\beta_P, \\
\frac{1}{a_F(P)} \frac{d}{d\tau_F} \left( \tilGa{\mu}{\alpha}{\beta} \right)_P & = & \left( \pdtilGa{\mu}{\alpha}{\beta}{\gamma} \right)_P (e_0)^\gamma_P.
\eea
The first equation holds since $(e_0)^\mu_P$ is the tangent unit vector of the central geodesic, while the second defines the conformal Hubble parameter of the CFC scale factor.  The third relation is a consequence of the stipulation that $(e_i)^\mu_P$ are parallel-transported along the central geodesic, and the fourth is obtained from the third directly through the chain rule. We eventually obtain
\bea
\label{eq:CFC-jac-time}
\frac{1}{a_F(P)} \frac{\partial x^\mu}{\partial \tau_F} & = & (e_0)^\mu_P + a_F \left[ \frac{\coH_F}{a_F} (e_i)^\mu - \Ga{\mu}{\alpha}{\beta} (e_0)^\alpha (e_i)^\beta  \right]_P x^i_F \nn\\
&& - a^2_F \left[ \frac{\coH_F}{a_F} \tilGa{\mu}{\alpha}{\beta} + \left( \frac12 \pdtilGa{\mu}{\alpha}{\beta}{\gamma} - \tilGa{\mu}{\alpha}{\sigma} \Ga{\sigma}{\beta}{\gamma} \right) (e_0)^\gamma \right]_P (e_i)^\alpha_P (e_j)^\beta_P \, x^i_F x^j_F \nn\\
&& + \mathcal{O}[(x^i_F)^3].
\eea 
The computation of $\partial x^\mu / \partial x^i_F$, by comparison, is more straightforward, since all quantities defined on the central geodesic do not depend on $x^i_F$. Hence we have
\bea
\label{eq:CFC-jac-space}
\frac{1}{a_F(P)} \frac{\partial x^\mu}{\partial x^l_F} & = & (e_l)^\mu_P - a_F(P) \left( \tilGa{\mu}{\alpha}{\beta} \right)_P (e_i)^\alpha_P (e_l)^\beta_P \, x^i_F \nn\\
&& - \frac16 a^2_F(P) \left[ \pdtilGa{\mu}{\alpha}{\beta}{\gamma} + 2 \pdtilGa{\mu}{\beta}{\gamma}{\alpha} - 2 \tilGa{\mu}{\gamma}{\sigma} \tilGa{\sigma}{\alpha}{\beta} - 4 \tilGa{\mu}{\alpha}{\sigma} \tilGa{\sigma}{\beta}{\gamma} \right]_P (e_i)^\alpha_P (e_j)^\beta_P (e_l)^\gamma_P\, x^i_F x^j_F \nn\\
&& + \mathcal{O}[(x^i_F)^3],
\eea
following directly from \refeq{CFC-coordinate-shift}. \refeq{CFC-jac-time} and \refeq{CFC-jac-space} can be compared to Eq.~(A9) and Eq.~(A10) of Ref.~\cite{Schmidt:2012nw}, respectively, which provide the corresponding relations for the standard FNC.  Instead of the Christoffel symbols $\Ga{\mu}{\alpha}{\beta}$, the conformal version appears $\tilGa{\mu}{\alpha}{\beta}$, in addition to $a_F$ and $\coH_F/a_F$.  Note that $a_F$ and $x_F^i$ always appear with the same power, so that only the combination $a_F\, x^i_F$ matters.

When \refeq{CFC-metric-conformal-transform} is computed order-by-order in $x^i_F$, we have to perform an expansion at several places. First, we have derived order-by-order expressions for the Jacobian matrix. Second, the conformal metric $\tilde g_{\mu\nu}\equiv a^{-2} g_{\mu\nu}$ has to be expanded about the CFC origin,
\bea
\label{eq:metric-shift}
\tilde g_{\mu\nu} (x^\lambda) = \left(\tilde g_{\mu\nu}\right)_P + \left( \partial_\alpha\, \tilde g_{\mu\nu} \right)_P \, \delta x^\alpha + \frac12 \left( \partial_\alpha \partial_\beta\, \tilde g_{\mu\nu} \right)_P\, \delta x^\alpha \delta x^\beta + \mathcal{O}[(\delta x^\mu)^3],
\eea
where $\delta x^\mu$ is given by \refeq{CFC-coordinate-shift}. Last but not least, we need to expand the factor $(a(Q)/a(P))^2$ as well. Since the global scale factor is only a function of the global time $\tau$, we first find the global time lag of a point at $(\tau_F,x^i_F)$ with respect to the CFC origin at $(\tau_F,0)$,
\bea
\delta \tau(x^i_F) \equiv \tau(Q) - \tau(P) = a_F(P) (e_i)^0_P \, x^i_F - \frac12 \left( \tilGa{0}{\alpha}{\beta} \right)_P [a_F(P)]^2 (e_i)^\alpha_P (e_j)^\beta_P \, x^i_F x^j_F + \mathcal{O}[(x^i_F)^3]. \nn\\
\eea
This induces a change in the global scale factor
\bea
\frac{a(Q)}{a(P)} = 1 + \coH(P) \delta \tau(x^i_F) + \frac12 \left( \coH'(P) + \coH^2(P) \right) [\delta \tau(x^i_F)]^2 + \mathcal{O}[\delta\tau^3(x^i_F)],
\eea
where $\coH(P),\,\coH'(P)$ are to be evaluated on the central geodesic.

An important simplification arises in \refeq{metric-shift} when considering a linearly perturbed FRW universe. The conformal metric $\tilde g_{\mu\nu}(x) = (a/a_F)^2 ( \eta_{\mu\nu} + h_{\mu\nu}(x))$ deviates from the flat metric only at first order in $h_{\mu\nu}$, and the zeroth-order piece $\eta_{\mu\nu}$ is a constant and does not contribute to derivatives. It is therefore sufficient to include the zeroth-order shift $\delta x^\mu = \delta^{\mu}_i\, x^i_F$, which is purely spatial [\refeq{CFC-coordinate-shift}].

\subsection*{Conformal flatness of CFC at $\mathcal{O}[x^i_F]$}
\label{app:CFC-metric-linear}

We now explicitly verify that the CFC metric is conformally flat, up to tidal corrections of $\mathcal{O}[(x^i_F)^2]$.  This is an independent but
equivalent calculation to the one in the CFC frame described in \refapp{CFC-metric-CFC}.  
The strategy is to check each of the three types of components of \refeq{CFC-metric-conformal-transform} --- $00$-component, $0i$-component, and $ij$-component. Algebra (e.g. partial derivatives) are carried out in global coordinates.

For the $00$-component, we expand to sufficient orders in $x^i_F$. Useful simplifications include $a^2(P) (\tilde g_{\mu\nu})_P = (g_{\mu\nu})_P$ and $a^2(P) (\tilde g_{\mu\nu})_P (e_\alpha)^\mu_P (e_\beta)^\nu_P = \eta_{\alpha\beta}$. In particular, we need expressions up to linear order in $x^i_F$,
\bea
\delta x^\mu = a_F(P) (e_i)^\mu_P x^i_F, \qquad \delta \tau = a_F(P) (e_i)^0 x^i_F.
\eea
We then have
\bea
&& 1 + (a_F)^{-2} g^F_{00} = 1 + \left[ \frac{a(P)}{a_F(P)} \frac{\partial x^\alpha}{\partial \tau_F}  \right] \left[ \frac{a(P)}{a_F(P)} \frac{\partial x^\beta}{\partial \tau_F}  \right] \left( \frac{a(\tau)}{a(P)} \right)^2 \left[ \left( \tilde g_{\alpha\beta} \right)_P + \left( \partial_\rho \tilde g_{\alpha\beta} \right)_P \delta x^\rho \right] \nn\\
& = &1 + \left[ (e_0)^\alpha_P + \left( \frac{\coH_F}{a_F} e^\alpha_i - \Ga{\alpha}{\mu}{\nu} e^\mu_0 e^\nu_i \right)_P a_F(P) x^i_F \right] \left[ (e_0)^\beta_P + \left( \frac{\coH_F}{a_F} e^\beta_j - \Ga{\beta}{\rho}{\sigma} e^\rho_0 e^\sigma_j \right)_P a_F(P) x^j_F \right] \nn\\
&& \times \left[ 1 + 2 \coH(P) (e_i)^0_P a_F(P) x^i_F \right] a^2(P) \left[ \left( \tilde g_{\alpha\beta} \right)_P + \left( \partial_\lambda \tilde g_{\alpha\beta} \right)_P (e_k)^\lambda_P a_F(P) x^k_F \right] \nn\\
& = & a_F(P) x^i_F \left[ 2 g_{\alpha\beta,P} \left( \frac{\coH_F}{a_F} e^\alpha_i e^\beta_0 - \Ga{\alpha}{\mu}{\nu} e^\mu_0 e^\nu_i e^\beta_0 \right)_P + a^2(P) \left( e^\alpha_0 e^\beta_0 e^\lambda_i \partial_\lambda \tilde g_{\alpha\beta} \right)_P + 2 \coH(P) \left( e^\alpha_0 e^\beta_0 e^0_i g_{\alpha\beta} \right)_P \right] \nn\\
& = & a_F(P) x^i_F \left[ - 2 \left( g_{\alpha\beta} \Ga{\alpha}{\mu}{\nu} e^\mu_0 e^\nu_i e^\beta_0 \right)_P + \left( e^\alpha_0 e^\beta_0 e^\lambda_i \partial_\lambda g_{\alpha\beta} \right)_P - 2 \left( e^\alpha_0 e^\beta_0 e^\lambda_i g_{\alpha\beta} \partial_\lambda \ln a \right)_P - 2 \coH(P) \left( e^0_i \right)_P \right]. \nn\\
\eea
The last two terms cancel because $\partial_\lambda \ln a = \coH \delta_{0\lambda}$. The first two terms also cancel because $2 g_{\alpha\beta} \Ga{\alpha}{\mu}{\nu} = \partial_\mu g_{\beta\nu} + \partial_\nu g_{\beta\mu} - \partial_\beta g_{\mu\nu}$.

We then check the $0i$-component. The conformal connection $\tilGa{\mu}{\alpha}{\beta}$ will be involved. We need the following expression at leading order in $x^i_F$,
\bea
\label{eq:ctensor-linear-order}
\Cten{\mu}{\alpha}{\beta} = - \frac{\coH_F}{a_F} \left( \delta^\mu_\alpha (e_0)_{P,\beta} + \delta^\mu_\beta (e_0)_{P,\alpha} - g_{\alpha\beta} (e_0)^\mu_P \right),
\eea
to compute $\tilGa{\mu}{\alpha}{\beta} = \Ga{\mu}{\alpha}{\beta} - \Cten{\mu}{\alpha}{\beta}$. We have
\bea
&& (a_F)^{-2} g^F_{0i} = \left[ \frac{a(P)}{a_F(P)} \frac{\partial x^\alpha}{\partial \tau_F}  \right] \left[ \frac{a(P)}{a_F(P)} \frac{\partial x^\beta}{\partial x^i_F}  \right] \left( \frac{a(\tau)}{a(P)} \right)^2 \left[ \left( \tilde g_{\alpha\beta} \right)_P + \left( \partial_\rho \tilde g_{\alpha\beta} \right)_P \delta x^\rho \right] \nn\\
& = & \left[ (e_0)^\alpha_P + \left( \frac{\coH_F}{a_F} e^\alpha_i - \Ga{\alpha}{\mu}{\nu} e^\mu_0 e^\nu_i \right)_P a_F(P) x^i_F \right] \left[ (e_i)^\beta_P - \left( \tilGa{\beta}{\rho}{\sigma} e^\rho_j e^\sigma_i \right)_P a_F(P) x^j_F \right] \nn\\
&& \times \left[ 1 + 2 \coH(P) (e_k)^0_P a_F(P) x^k_F \right] a^2(P) \left[ \left( \tilde g_{\alpha\beta} \right)_P + \left( \partial_\lambda \tilde g_{\alpha\beta} \right)_P (e_l)^\lambda_P a_F(P) x^l_F \right] \nn\\
& = & a_F(P) x^j_F \left[ \frac{\coH_F}{a_F} \delta_{ij} - \left( \Ga{\alpha}{\mu}{\nu} e^\mu_0 e^\nu_j e_{i,\alpha} \right)_P - \left( \tilGa{\beta}{\mu}{\nu} e^\mu_j e^\nu_i e_{0,\beta} \right)_P + a^2(P) \left( e^\alpha_0 e^\beta_i e^\lambda_j \partial_\lambda \tilde g_{\alpha\beta} \right)_P \right] \nn\\
& = &  a_F(P) x^j_F \left[ \frac{\coH_F}{a_F} \delta_{ij} - \left( \Ga{\alpha}{\mu}{\nu} e^\mu_0 e^\nu_j e_{i,\alpha} \right)_P - \left( \Ga{\beta}{\mu}{\nu} e^\mu_j e^\nu_i e_{0,\beta} \right)_P + \left( \Cten{\beta}{\mu}{\nu} e^\mu_j e^\nu_i e_{0,\beta} \right)_P \right.\nn\\
&& \left. + \left( e^\alpha_0 e^\beta_i e^\lambda_j \partial_\lambda g_{\alpha\beta} \right)_P \right].
\eea
The first term will cancel with the fourth term by \refeq{ctensor-linear-order}. The second, the third and the last term add up to zero after expressing the Christoffel symbols in terms of $\partial_\lambda g_{\alpha\beta}$. Therefore, the whole expression vanishes at $\mathcal{O}[x^i_F]$.

Finally, we check the $ij$-component. We find
\bea
&& - \delta_{ij} + (a_F)^{-2} g^F_{ij} = -\delta_{ij} + \left[ \frac{a(P)}{a_F(P)} \frac{\partial x^\alpha}{\partial x^i_F}  \right] \left[ \frac{a(P)}{a_F(P)} \frac{\partial x^\beta}{\partial x^j_F}  \right] \left( \frac{a(\tau)}{a(P)} \right)^2 \left[ \left( \tilde g_{\alpha\beta} \right)_P + \left( \partial_\rho \tilde g_{\alpha\beta} \right)_P \delta x^\rho \right] \nn\\
& = & - \delta_{ij} + \left[ (e_i)^\alpha_P - \left( \tilGa{\alpha}{\mu}{\nu} e^\mu_k e^\nu_i \right)_P a_F(P) x^k_F \right] \left[ (e_j)^\beta_P - \left( \tilGa{\beta}{\rho}{\sigma} e^\rho_l e^\sigma_j \right)_P a_F(P) x^l_F \right] \nn\\
&& \times \left[ 1 + 2 \coH(P) (e_k)^0_P a_F(P) x^k_F \right] a^2(P) \left[ \left( \tilde g_{\alpha\beta} \right)_P + \left( \partial_\lambda \tilde g_{\alpha\beta} \right)_P (e_l)^\lambda_P a_F(P) x^l_F \right] \nn\\
& = & a_F(P) x^k_F \left[ - \left( \tilGa{\alpha}{\mu}{\nu} e^\mu_k e^\nu_i e_{j,\alpha}\right)_P - \left( \tilGa{\beta}{\mu}{\nu} e^\mu_k e^\nu_j e_{i,\beta}\right)_P   + 2\coH(P) (e_k)^0_P \delta_{ij} + a^2(P) \left( e^\alpha_i e^\beta_j e^\lambda_k \partial_\lambda \tilde g_{\alpha\beta} \right)_P \right] \nn\\
& = & a_F(P) x^k_F \left[ - \left( \tilGa{\alpha}{\mu}{\nu} e^\mu_k e^\nu_i e_{j,\alpha}\right)_P - \left( \tilGa{\beta}{\mu}{\nu} e^\mu_k e^\nu_j e_{i,\beta}\right)_P + 2\coH(P) (e_k)^0_P \delta_{ij} \right. \nn\\
&& \left. + \left( e^\alpha_i e^\beta_j e^\lambda_k \partial_\lambda g_{\alpha\beta} \right)_P - 2 \delta_{ij} \left( e^\lambda_k \partial_\lambda \ln a \right)_P  \right].
\eea
The third term cancels with the last term by $\partial_\lambda \ln a = \coH \delta_{0\lambda}$. The remaining three terms sum up to zero. In particular, since $\Cten{\alpha}{\mu}{\nu}$ vanishes when contracting with three spatial tetrad vectors, the extra pieces in the conformal connections do not contribute.

We therefore have proved \refeq{CFC-requirement}. The proof does not assume that the global metric $g_{\mu\nu}$ is parameterised in small perturbations, and therefore holds at all orders in perturbation fields (not to be confused with expansion in $x^i_F$).

\section{Structure of CFC metric corrections}
\label{app:CFC-metric-CFC}

This appendix supplements algebraic details underlying the discussion of \refsec{CFC-metric}.  That is, we show that the metric in CFC is of the form \refeq{CFC-requirement}, and derive the explicit expression for the $\O(x_F^2)$ corrections.  We also discuss the higher order terms that appear at $x_F^3$ and higher.  
Throughout we will only deal with quantities in CFC, nevertheless we will keep the superscript $F$ here.  

The CFC are defined by, first, that \refeq{eq-CFC-zeroth} holds for all times along the central geodesic.  Second, we require that $(\tilde \Gamma^F)^\mu_{\alpha\beta} |_{\rm G}=0$, which is equivalent to \refeq{eq-CFC-first-deriv}.  
To show this, we consider again the curve \refeq{eq-CFC-spatial-geo-par}. It describes a geodesic with respect to the conformal metric $\tilde g$, and it is linear in the affine parameter $\lambda$. In CFC, the tangent vector is $(0,\beta^i)$, and the conformal geodesic equation \refeq{conformal-geodesic-eq} reduces to
\bea
\left. (\tilde \Gamma^F )^\mu_{ij} \right|_P \,\beta^i \beta^j = 0,
\eea
which implies $(\tilde \Gamma^F)^\mu_{ij} |_P=0$ because $\beta^i$ is arbitrary. To examine the remaining symbols $(\tilde \Gamma^F)^\mu_{0 \alpha}$, we use that $(e_\alpha)^\mu=a^{-1}_F\,\delta^\mu_\alpha$ are parallel-transported along the central geodesic. In CFC, $(e_0)^\nu \nabla_\nu (e_\alpha)^\mu = 0$ reduces to
\bea
a^{-1}_F \partial_0 \left( a^{-1}_F \delta^\mu_\alpha \right) + a^{-2}_F \left(\Gamma^F \right)^\mu_{\nu\rho} \delta^\nu_\alpha \delta^\rho_0 = 0,
\eea
which becomes
\bea
\left. \left( \Gamma^F \right)^\mu_{0 \alpha} \right|_P = \left. \left( \ln a_F \right)' \right|_P\,\delta^\mu_\alpha,
\eea
where throughout this appendix a prime denotes a derivative with respect to $\tau_F$. We then merely have to relate the ordinary symbol $\Gamma^F$ to the conformal one $\tilde \Gamma^F$ (components evaluated in CFC) using \refeq{tilGa} and \refeq{Ctensor}. The gradient of $a_F$ is required to be tangent to the central geodesic, so we have $\nabla_\mu \ln a_F = \delta^0_\mu\, (\ln a_F)'$ and
\bea
\left. \left( C^F \right)^\mu_{0\alpha} \right|_P = \left( \ln a_F \right)' \delta^\mu_\alpha.
\eea
Combining this with the result for $(\Gamma^F)^\mu_{0\alpha}$ leads to
\bea
\left. ( \tilde \Gamma^F )^\mu_{0 \alpha} \right|_P = 0.
\eea 
This guarantees that there are no terms linear in $x^i_F$ in the CFC metric.

To derive the $\mathcal{O}[(x^i_F)^2]$ terms, we directly apply the strategy of Ref.~\cite{Manasse:1963zz}. For FNC, the authors deal with Christoffel symbols and Riemann tensor with respect to the physical metric. For CFC, all results carry through, as long as we instead deal with quantities defined with respect to the conformally-related metric. To be specific, we want to relate $\partial_\alpha\partial_\beta\,\tilde g^F_{\mu\nu}$ to $\tilde R^F_{\mu\nu\rho\sigma}$ in CFC at some point along the central geodesic. Again, since $(\tilde \Gamma^F)^\alpha_{\mu\nu}$ always vanishes there, we have
\bea
\label{eq:D5}
\left. \partial_0 (\,\tilde \Gamma^F )^\alpha_{\mu\nu} \right|_P = 0.
\eea
Besides, we have in CFC
\bea
\label{eq:D6}
\left. \partial_\nu (\,\tilde \Gamma^F )^\alpha_{\mu 0} \right|_P = \left. (\tilde R^F)^\alpha{}_{\mu\nu 0} \right|_P.
\eea
The remaining derivatives $\partial_k (\tilde \Gamma^F)^\alpha_{ij}$ involve three spatial indices downstairs. The derivation parallels the one detailed in Ref.~\cite{Manasse:1963zz}, which we do not repeat here but only outline the idea. We can consider the geodesic deviaton equation (with respect to $\tilde g$) applied to the tangent vector $n_i \sim \partial/\partial x^i_F$ of the spatial geodesic (see also \refsec{CFC-lagrange-eulerian}; note that unlike the central geodesic itself, the spatial CFC coordinate lines are geodesic with respect to $\tilde g$). The geodesic deviation equation has the schematic form $\nabla \nabla n + \tilde R^F n = 0$, where the second covariant derivative reduces to the spatial derivative of $\tilde \Gamma^F$ and therefore leads to a formula $\partial\, \tilde \Gamma^F = \tilde R^F$. The final result reads
\bea
\label{eq:D7}
\left. \partial_k (\,\tilde \Gamma^F )^\alpha_{ij} \right|_P = - \frac13 \left[ (\tilde R^F)^\alpha{}_{ijk} + (\tilde R^F)^\alpha{}_{jik} \right]_P.
\eea
Combining \refeq{eq-CFC-zeroth} with \refeqs{D5}{D7}, we obtain double spatial derivatives of $\tilde g$,
\bea
\label{eq:eq-D9}
\left. \partial_i \partial_j\, \tilde g^F_{00} \right|_P & = & - 2 \left. \tilde R^F_{0i0j} \right|_P, \\
\left. \partial_i \partial_j\, \tilde g^F_{0k} \right|_P & = & - \frac23 \left. \left( \tilde R^F_{0ikj} + \tilde R^F_{0jki} \right) \right|_P, \\
\left. \partial_i \partial_j\, \tilde g^F_{kl} \right|_P & = & - \frac13 \left. \left( \tilde R^F_{ikjl} + \tilde R^F_{jlik} \right) \right|_P.
\eea
These results imply \refeqs{gF-00}{gF-ij}.

We now discuss what the next, higher order contribution
to the metric at $\mathcal{O}[(x^i_F)^3]$ looks like. On the central geodesic $G$, we examine
\be
\left. \partial_i \partial_j \partial_k \,\tilde g^F_{\mu\nu} \right|_P\,,
\ee 
where $P$ denotes an arbitrary point on G which we keep fixed.  
Note that we do not need to consider time derivatives in CFC,
since these can be consistently taken within the leading CFC frame metric.  
(We hereafter drop the super/subscript $F$ as we will deal
exclusively with quantities in CFC). We have
\be
\tilde\Gamma^\mu_{\alpha\beta} = \tilde g^{\mu\nu} \left( 2\partial_{(\alpha} \tilde g_{\nu\beta)} - \partial_\nu \tilde g_{\alpha\beta} \right)\,.
\ee
Thus, taking two spatial derivatives and evaluating on $G$ yields
\be
\partial_i \partial_j \tilde\Gamma^\mu_{\alpha\beta} \Big|_P
= \eta^{\mu\nu} \partial_i\partial_j\left( 2\partial_{(\alpha} \tilde g_{\nu\beta)} - \partial_\nu \tilde g_{\alpha\beta} \right) \Big|_P
+ \O\left[ (\partial \tilde g^{\mu\nu}) (\partial\partial \tilde g_{\gamma\delta}),\  
(\partial\partial \tilde g^{\mu\nu}) (\partial \tilde g_{\gamma\delta}) \right]\Big|_P\,.
\ee
Only the first term, involving three derivatives of the metric, is
nonzero on $G$.  In order to extract the triple spatial derivative terms
of the metric (as explained above, we already know all the other ones),
we need to consider $\partial_i\partial_j \tilde\Gamma^\mu_{0k},\,\partial_i\partial_j \tilde\Gamma^\mu_{kl}$. As shown in \cite{Manasse:1963zz}, since $\tilde\Gamma$ vanishes on $G$ we have
\be
\partial_j \tilde\Gamma^\mu_{0k}\Big|_P = -\tilde R^\mu_{\  k0j} \Big|_P\,.
\ee
We then obtain
\be
\partial_i\partial_j \tilde\Gamma^\mu_{0k}\Big|_P = - \partial_i \tilde R^\mu_{\  k0j} \Big|_P\,,
\ee
since the correction from the $(\tilde \Gamma)^2$ term in $\tilde R$ vanishes
even after taking a derivative.  The spatial derivatives of $\tilde\Gamma^\mu_{kl}$ can be obtained by taking another derivative with respect to the tangent $T$ along the spatial geodesic of the geodesic deviation equation.  Again neglecting terms of order $\tilde\Gamma \,\partial\tilde\Gamma$ which vanish on $G$, this yields simply the ordinary CFC spatial derivative of the geodesic deviation equation written in CFC, Eq.~(64c) of \cite{Manasse:1963zz}. Schematically,
\be
\partial_i \partial_j \tilde\Gamma^\mu_{kl} \Big|_P = \partial_i \left[ \tilde R^\mu_{\  klj} + {\rm perm.} \right] \Big|_P\,.
\ee
To conclude, the $\mathcal{O}[(x^i_F)^3]$ corrections to the CFC frame conformal metric scale as
\be
\left( \partial^F_i \tilde R^F_{\mu j \nu k}\right)_P\, x_F^i x_F^j x_F^k\,,
\ee
i.e. they are suppressed by \emph{the spatial derivative in CFC of $\tilde R^F$} multiplied by $x_F$.  At order $(x_F^i)^4$, we similarly have terms of order
$\partial\partial\tilde R^F$, but we also in general obtain terms of order
$(\tilde R^F)^2$.  Thus, we are performing an expansion in powers of 
$\tilde R^F$ and its spatial derivatives, with appropriate powers of $x_F$.  

\section{Inverse spatial derivative in CFC}
\label{app:CFC-inv-spatial-deriv}

In this Appendix we discuss how inverse spatial derivatives are to
be performed in CFC, which break translation invariance.  
We show that it amounts to the following.  
One formally writes $\partial = \partial_S + \partial_L$ and then Taylor expands in powers of $\partial_L/\partial_S$, subject to the rule that $\partial_S$'s only act on short-wavelength perturbations and $\partial_L$'s only on $x^i_F$. The algebra is then straightforward because no negative power of $\partial_L$ exists.  Since we consistently drop any $\O([x^i]^3)$ and higher corrections to 
the CFC metric, this expansion is truncated at $\partial_L^2$.

In CFC, at $\mathcal{O}[(x^i)^n]$, a generic term on which one desires to act with (inverse-)spatial derivative looks like
\bea
\label{eq:eq-generic-term}
f(x^i,\tau) F(\tau)\, x^{i_1} x^{i_2} \cdots x^{i_n},
\eea
where $f(x^i,\tau)$ only depends on the short modes, and $F(\tau)$ only depends on the long modes.  Note that $F(\tau)$ only depends on time by definition, since it is evaluated on the central geodesic.   We have suppressed the spatial indices of $f$ and $F$ for clarity.  To simplify our notation further, we suppress the time dependence hereafter. 

\refeq{eq-generic-term} does not have a well-defined Fourier transform, since
it is not bounded at infinity.  In order to treat this, we let $F(\tau)$ have
a smooth, very large-scale shape in real space, corresponding to very narrow
support around $0$ in $k$ space.  Thus, in $k$-space we multiply $F(\tau)$
with 
\bea
\frac{C}{\Lambda^3} \exp\left( - \frac{k^2}{\Lambda^2} \right),
\eea
where $C$ is a normalization constant and $\Lambda$ is the cutoff in Fourier
space.  Apart from factors of order one, $\Lambda$ is nothing else than the coarse-graining scale introduced
in \refsec{EE}.  If we take $\Lambda \rightarrow 0$, this factor becomes $(2\pi)^3 \delta_D^{(3)}(\vec k)$, i.e. the Fourier transform of unity as desired. If we use $~\widetilde{}~$ to represent the Fourier transform of a field in real space, the Fourier transform of \refeq{eq-generic-term} is a convolution
\bea
\left[ \widetilde{f(x^i,\tau) F(\tau)\, x^{i_1} x^{i_2} \cdots x^{i_n}}
\right] (\vec k) & = & \int \frac{d^3 k_L}{(2\pi)^3} \, \frac{C}{\Lambda^3}\,\tilde f(\vec k_S) \, F \frac{-i \partial}{\partial k^{i_1}_L} \frac{-i \partial}{\partial k^{i_2}_L} \cdots \frac{-i \partial}{\partial k^{i_n}_L} \exp\left( - \frac{k^2_L}{\Lambda^2} \right),\quad\quad
\eea
where $\vec k_S = \vec k - \vec k_L$. The number of $-i\partial/\partial k_L$ operators acting on the exponential counts the order in $x^i$.

We now use $\partial^{-2}$ as an example to demonstrate how inverse, and therefore non-local, spatial derivatives can be computed for a power expansion in $x^i$. It amounts to power expanding the operator in Fourier space
\bea
\partial^{-2} & \longrightarrow & \frac{-1}{k^2} = - \frac{1}{k^2_S} \left[ 1 - 2 \frac{k_{S,i}}{k^2_S} k^i_L + \left( - \frac{1}{k^2_S} \delta_{ij} + 4 \frac{k_{S,i} k_{S,j}}{(k^2_S)^2} \right) k^i_L k^j_L + \mathcal{O}[(k_L)^3] \right],
\eea 
so that $\vec k_L$ only appears in the numerator. Powers of $\vec k_S$ then trivially translate into (inverse-)spatial derivatives of the short mode $f$.

For each power of $k_L$, we can commute it with $-i\partial/\partial k_L$'s until it appears nearest to the exponential; but then this term vanishes in the limit $\Lambda \rightarrow 0$. To convince ourselves, we have
\bea
k^i_L \exp\left( - \frac{k^2_L}{\Lambda^2} \right) & = & \frac{i\Lambda^2}{2} \frac{-i\partial}{\partial k_{L,i}} \exp\left( - \frac{k^2_L}{\Lambda^2} \right).
\eea
In real space, this seemingly raises the power of $x^i$ by one; however, because of the two extra powers of $\Lambda$, it approaches zero once the limit $\Lambda\rightarrow 0$ is taken in the end. Therefore, the surviving terms are those generated from the commutation relation $\comm{k^i_L}{\partial/\partial k^j_L} = - \delta^i{}_j$, which in real space translates into $\comm{x^i}{\partial_j} = - \delta^i{}_j$, identical to the action of $\partial_i$ on powers of $x^i$. The net result of our procedure boils down to writing $\partial^i = \partial^i_S + \partial^i_L$, Taylor-expanding in $\partial_L/\partial_S$, and following the rule that $\partial^i_S$'s only act on $f$, and $\partial^i_L$'s act on powers of $x^i$. For instance, we have for the non-local operator $\partial^i \partial^j/\partial^2$ (which is used in our calculation for this work)
\bea
\label{eq:longi-op-taylor}
\frac{\partial^i \partial^j}{\partial^2} & = & \frac{\partial^i_S \partial^j_S}{\partial^2_S} + \left( 2 \frac{\partial^{(i}_S \delta^{j)}_a }{\partial_S^2} - \frac{2 \partial^i_S \partial^j_S}{(\partial^2_S)^2} \partial_{S,k}  \right) \partial^k_L \nn\\
&& + \left[ \frac{\delta^i_k \delta^j_l}{\partial^2_S} - \frac{ \partial^{(i}_S \partial^{j)}_S \delta_{kl} + 4 \partial^{(i}_S \delta^{j)}_k \partial_{S,l} }{(\partial^2_S)^2} + 4 \frac{\partial^{(i}_S \partial^{j)}_S \partial_{S,k} \partial_{S,l}}{(\partial^2_S)^3} \right] \partial^k_L \partial^l_L + \mathcal{O}\left[\left(\frac{\partial_L}{\partial_S}\right)^3\right].\quad
\eea 
The power expansion in $\partial_L/\partial_S$ truncates for local operators, but is in general an infinite series for non-local operators. Therefore, non-local operators in principle mix up all orders of $x^i$ in CFC.  
In practice, the series is truncated at finite order. The reason is that each additional power of $\partial_L$ comes with an inverse $\partial_S$, and is therefore suppressed by one more power of $k_L/k_S$. In our computation, we drop terms that are proportional to a net negative power of $\partial_S$ acting on the short-scale potential. These necessarily come with three powers of $\partial_L$ and thus require the CFC metric beyond quadratic order. Therefore, only a finite number of terms in the power expansion need to be considered.

\section{Second-order anisotropic stress in CFC}
\label{app:CFC-2nd-aniso-stress}

We start with the second order part of the spatial-spatial Einstein equation $G^i{}_j = 8\pi G T^i{}_j$,
\bea
&& \left[ 2 \psi^{(2)}{}'' + \coH \left( 4 \psi^{(2)}{}' + 2 \phi^{(2)}{}' - \partial^2 \left( \psi^{(2)} - \phi^{(2)} \right) \right) \right] \delta^i{}_j + \partial^i \partial_j \left( \psi^{(2)} - \phi^{(2)} \right) \nn\\
&& + \mathcal{O}[(\partial \partial \phi^{(1)}) h^F_{\mu\nu}] + \mathcal{O}[(\partial \phi^{(1)}) \left( \partial h^F_{\mu\nu} \right)] + \mathcal{O}[\phi^{(1)} \left( \partial \partial h^F_{\mu\nu} \right)] \, = \, 8 \pi G\, v^{(1)i} v^{(1)}_j.
\eea
On the left hand side, we first of all collect second-order short-scale perturbations. In addition, from an expansion of $G^i{}_j$ at second order, we have a bunch of quadratic terms involving linear short modes (or its derivatives) multiplied by the CFC tidal metric $h^F_{\mu\nu}$ (given in \refeqs{tensor-tidal-force-00}{tensor-tidal-force-ij}) or their derivatives, which we do not explicitly spell out here. Since the CFC metric correction $h^F_{\mu\nu}$ is truncated at $\mathcal{O}[(x^i)^2]$, these terms are at most $\mathcal{O}[(x^i)^2]$. On the right hand side, we have a term quadratic in $v^{(1)}_i$ from $T^i{}_j$. We only have to insert the solution for $v^{(1)}_i$ up to $\mathcal{O}[x^i]$ (\refeq{linsol}), because the higher orders are generated by CFC metric correction $h^F_{\mu\nu}$ beyond the quadratic order, which is subleading in $k_L/k_S$. 

The longitudinal scalar part of this tensorial equation is obtained by acting upon it with the non-local operator
\bea
\frac{\partial_i \partial^j}{\partial^2} - \frac13 \delta_i{}^j,
\eea 
as defined in CFC by the Taylor expansion \refeq{longi-op-taylor}, and then evaluating at $x^i=0$. According to the general discussion in \refapp{CFC-inv-spatial-deriv}, it is sufficient to expand to $\mathcal{O}[(\partial_L)^2]$ in \refeq{longi-op-taylor}. 

With the prescription of \refapp{CFC-inv-spatial-deriv}, straightforward but tedious algebra can be carried out. However, even at $x^i=0$, more terms can be dropped. Indeed, we learn from \refapp{CFC-inv-spatial-deriv} that, beyond leading order in $\partial_L/\partial_S$, eliminating one power of $x^i$ by one power of $\partial_L$ occurs at the expense of introducing one more power of $\partial_S$ in the denominator, which brings about an additional suppression by a factor $k_L/k_S$. Simple power counting suggests that to retain leading terms in $k_L/k_S$, {\it we can drop all terms having a net negative power of $\partial_S$ acting on the short mode}.

The end result on the central geodesic reads
\bea
\frac23 \partial^2 \Gamma^{(2)}  - \left( \gamma''_{ij} + \partial^2 \gamma_{ij} \right) \left( \frac{\partial^i \partial^j}{\partial^2} \phi^{(1)} \right)  & = & 2 \coH \gamma_{ij}' \left( \frac{\partial^i \partial^j}{\partial^2} \phi^{(1)} \right).
\eea
The left hand side comes from $G^i{}_j$ while the right hand side comes from $T^i{}_j$. This leads to the solution for the second-order ``effective'' anisotropic stress \refeq{Gamma-2nd-res}.

\section{Comparison with previous results}
\label{app:CFC-compare-Dai}

In Ref.~\cite{Dai:2013kra}, the density perturbation to second order was computed in global coordinates,
\bea
\label{eq:eq-E.1-deltaG}
\delta_G & = & \left( -2 + \frac{2}{3\coH^2} \partial^2 \right) \phi_{\rm ini} - \frac{2}{3\coH^2} \left[ \mathcal{T}_\gamma(k_L) + \mathcal{S}(k_L) \right] \gamma^{ij}_{\rm ini} \left( \partial_i \partial_j \phi_{\rm ini} \right) \nn\\
&& + \frac{2}{\coH^2} \mathcal{S}_N (k_L)\left( \partial^2 \gamma^{ij}_{\rm ini} \right) \left( \frac{\partial_i \partial_j}{\partial^2} \phi_{\rm ini} \right)  + 2 \gamma^{ij}_{\rm ini} \left( \frac{\partial_i \partial_j}{\partial^2} \phi_{\rm ini} \right) \left[ \frac{1}{\coH} \frac{d\mathcal{S}(k_L)}{d\tau} + \mathcal{S}(k_L) \right].
\eea
Only the coefficient of the second term $\sim \partial_i \partial_j \phi_{\rm ini}$ does not agree exactly with the CFC result \refeq{eq-7.31}. The discrepancy is due to the fact that although results in both frames depend on the same (and physical) {\it initial} potential profile $\phi_{\rm ini}$, $\partial^2 \phi_{\rm ini}$ are in general different {\it scalar functions} for different parameterisation of the spatial coordinates.  This is because $\partial^2 \equiv \delta^{ij} \partial_i \partial_j$ is not coordinate invariant.

If the spatial origin of both frames is chosen to coincide, then to leading order, {\it on the initial spatial slice}, the global parameterisation can be converted into the CFC parameterisation (proper distance) via a shift
\bea
x^i & \longrightarrow & x^i - \frac12 \gamma^{ij}_{\rm ini}\, x_j\,.
\eea
If we label quantities in CFC by ${}^F$ and those in the global frame by ${}^G$, then on the initial slice we have (note that $\gamma^{ij}_{\rm ini}$ refers to the value at the origin and therefore should be regarded as a constant)
\bea
\phi^G_{\rm ini}\left( x^i \right) = \phi^F_{\rm ini}\left(x^i - \frac12 \gamma^{ij}_{\rm ini} x_j \right) =  \phi^F_{\rm ini}\left(x^i\right) + \left( \partial_i \phi_{\rm ini} \right)^F(x^i)\, \frac12 \gamma^{ij}_{\rm ini}\, x_j + \mathcal{O}[(\gamma_{\rm ini})^2].
\eea
Taking the spatial derivative twice, and then evaluating at $x^i=0$, we find
\bea
\left( \partial^2 \phi_{\rm ini} \right)^G & = & \left( \partial^2 \phi_{\rm ini} \right)^F - \gamma^{ij}_{\rm ini} \left( \partial_i \partial_j \phi_{\rm ini} \right)^F.
\eea
At the order of perturbation we pursue, this only applies to the first term (linear solution) of \refeq{eq-E.1-deltaG} and yields an additional term at second order, so that
\bea
\delta_G & \longrightarrow & \left( -2 + \frac{2}{3\coH^2} \partial^2 \right) \phi_{\rm ini} - \frac{2}{3\coH^2} \left[ \mathcal{T}_\gamma(k_L) - 1 + \mathcal{S}(k_L) \right] \gamma^{ij}_{\rm ini} \left( \partial_i \partial_j \phi_{\rm ini} \right) \nn\\
&& + \frac{2}{\coH^2} \mathcal{S}_N (k_L)\left( \partial^2 \gamma^{ij}_{\rm ini} \right) \left( \frac{\partial_i \partial_j}{\partial^2} \phi_{\rm ini} \right)  + 2 \gamma^{ij}_{\rm ini} \left( \frac{\partial_i \partial_j}{\partial^2} \phi_{\rm ini} \right) \left[ \frac{1}{\coH} \frac{d\mathcal{S}(k_L)}{d\tau} + \mathcal{S}(k_L) \right].
\eea
Given \refeq{eq-SN-S}, this is in complete agreement with \refeq{eq-7.31}.
 


\end{document}